%% file: main.tex
\documentclass[]{IEEEtran}

\usepackage{cite}
\usepackage{framed,multirow}
\usepackage{xargs}
\usepackage{booktabs}
\usepackage{amsmath}
\usepackage{color, colortbl}
\usepackage[pdftex,dvipsnames]{xcolor}
\usepackage[utf8]{inputenc}
\usepackage[T1]{fontenc}
\usepackage[pdfencoding=auto,bookmarks=false,hidelinks]{hyperref}
\usepackage{tikz}
\usepackage{mathtools}
\usepackage{textcomp}
\usepackage[font=small]{caption}
\usepackage{subcaption}
\usepackage{csvsimple}
\usepackage{amsmath}
\usepackage{amssymb}
\usepackage{siunitx}
\usepackage{algorithmicx}
\usepackage{algpseudocode}
\usepackage{setspace}
\usepackage{adjustbox}
\usepackage{array}
\usepackage{multicol}
\usepackage{gensymb}
\usepackage{wrapfig}
\usepackage{dblfloatfix}    % To enable figures at the bottom of page
\newcommand{\subparagraph}{}
\usepackage{titlesec}
\titlespacing*{\subsection}
{1ex}{1ex plus 1ex minus .2ex}{0.2ex plus .2ex}

\usepackage{pgfplots}
\pgfplotsset{compat=1.14,
/pgfplots/ybar legend/.style={
/pgfplots/legend image code/.code={
\draw [##1,/tikz/.cd,bar width=4pt,yshift=-0.2em,bar shift=0pt]
plot coordinates {(0cm,0.8em)};
}
}}
\usepgfplotslibrary{fillbetween}
\usetikzlibrary{arrows, positioning, shapes.geometric}

\newcommand{\norm}[1]{\left\lVert#1\right\rVert}
\newcommand{\ra}[1]{\renewcommand{\arraystretch}{#1}}

\definecolor{newcolor}{rgb}{.8,.349,.1}
\definecolor{incomp_color}{RGB}{255,128,245}
\definecolor{comp_color}{RGB}{69,255,81}

\usepackage{scalerel}
\usepackage{tikz}
\usetikzlibrary{svg.path}

\definecolor{orcidlogocol}{HTML}{A6CE39}
\tikzset{
  orcidlogo/.pic={
    \fill[orcidlogocol] svg{M256,128c0,70.7-57.3,128-128,128C57.3,256,0,198.7,0,128C0,57.3,57.3,0,128,0C198.7,0,256,57.3,256,128z};
    \fill[white] svg{M86.3,186.2H70.9V79.1h15.4v48.4V186.2z}
                 svg{M108.9,79.1h41.6c39.6,0,57,28.3,57,53.6c0,27.5-21.5,53.6-56.8,53.6h-41.8V79.1z M124.3,172.4h24.5c34.9,0,42.9-26.5,42.9-39.7c0-21.5-13.7-39.7-43.7-39.7h-23.7V172.4z}
                 svg{M88.7,56.8c0,5.5-4.5,10.1-10.1,10.1c-5.6,0-10.1-4.6-10.1-10.1c0-5.6,4.5-10.1,10.1-10.1C84.2,46.7,88.7,51.3,88.7,56.8z};
  }
}

\newcommand\orcidicon[1]{\href{https://orcid.org/#1}{\mbox{\scalerel*{
\begin{tikzpicture}[yscale=-1,transform shape]
\pic{orcidlogo};
\end{tikzpicture}
}{A}}}}

\newlength{\adaptivefigwidth}
\begin{document}
\bstctlcite{IEEEexample:BSTcontrol} % Setup Citation Style

% Reduce space above/below equations
\abovedisplayshortskip=3pt
\belowdisplayshortskip=5pt
\abovedisplayskip=5pt
\belowdisplayskip=5pt

\title{Learning Multiparametric Biomarkers for Assessing MR-Guided Focused Ultrasound Treatment of Malignant Tumors}%

\author{Blake~E.~Zimmerman\,\orcidicon{0000-0003-1769-7943},
        Sara~Johnson\,\orcidicon{0000-0001-8796-1549},
        Henrik~Od\'een\,\orcidicon{0000-0003-2055-9795},
        Jill~Shea\,\orcidicon{0000-0002-8644-7772},
        Markus~D.~Foote\,\orcidicon{0000-0002-5170-1937},
        Nicole~Winkler, 
        Sarang~C.~Joshi, 
        Allison~Payne\,\orcidicon{0000-0002-7724-5001}%
\thanks{Manuscript compiled \today.
This work was supported by the National Institutes of Health [5R37CA224141, S10OD018482, 1R03EB026132]}% <-this % stops a space
\thanks{B. E. Zimmerman, M. D. Foote, and S. C. Joshi are with the Scientific Computing and Imaging Institute, and also with the Department of Biomedical Engineering, University of Utah, 
Salt Lake City, UT 84112 USA.}% <-this % stops a space
\thanks{S. Johnson is with the Department of Biomedical Engineering, University of Utah, 
Salt Lake City, UT 84112 USA.}% <- this stops a space
\thanks{B. E. Zimmerman, S. Johnson, H. Od\'een, and A. Payne are with the Utah Center for Advanced Imaging Research, University of Utah, Salt Lake City, UT, USA.}% <- this stops a space
\thanks{S. Johnson, H. Od\'een, and A. Payne and N. Winkler are with the Department of Radiology and Imaging Sciences, University of Utah, Salt Lake City, UT, USA}%
\thanks{J. Shea is with the Department of Surgery, University of Utah, Salt Lake City, UT 84112 USA.}% <-this % stops a space
%\thanks{Color versions of one or more of the figures in this article are available online at http://ieeexplore.ieee.org.}% <-this % stops a space
\thanks{This is the author's version of the accepted paper. The publisher's copy may be accessed from IEEE by the DOI \href{https://doi.org/10.1109/TBME.2020.3024826}{10.1109/TBME.2020.3024826}\vspace{0.45cm}}% <-this % stops a space
}% <-this % stops a space and closes the author/thanks block

\markboth{Zimmerman \MakeLowercase{\textit{et al.}}: Learning Multiparametric Biomarkers for Assessing MR-Guided Focused Ultrasound Treatments}
{Zimmerman \MakeLowercase{\textit{et al.}}: Learning Multiparametric Biomarkers for Assessing MR-Guided Focused Ultrasound Treatments}

\maketitle
\IEEEpubid{\begin{minipage}{1.08\textwidth}\centering\scriptsize\copyright~2020 IEEE. 
        Personal use of this material is permitted. 
        Permission from IEEE must be obtained for all other uses, in any current or 
        future media, including reprinting/republishing this material for advertising 
        or promotional purposes, creating new collective works, for resale or 
        redistribution to servers or lists, or reuse of any copyrighted component 
        of this work in other works.\vspace*{-1.0cm}\end{minipage}}

\input{abstract}

\begin{IEEEkeywords}
machine learning, ablation therapy, deformable image registration, convolutional neural network
\end{IEEEkeywords}

\input{intro.tex}
\input{studydesign.tex}
\input{registration.tex}

\input{validation.tex}
\input{network.tex}

\input{results.tex}
\input{discussion.tex}

\input{acknowledgements.tex}

\bibliographystyle{IEEEtran}
\bibliography{refs}

\end{document}

%% file: abstract.tex
\begin{abstract}
Noninvasive MR-guided focused ultrasound (MRgFUS) treatments are promising alternatives to the surgical removal of malignant tumors. 
A significant challenge is assessing the viability of treated tissue during and immediately after MRgFUS procedures.
Current clinical assessment uses the nonperfused volume (NPV) biomarker immediately after treatment from contrast-enhanced MRI. 
The NPV has variable accuracy, and the use of contrast agent prevents continuing MRgFUS treatment if tumor coverage is inadequate.
This work presents a novel, noncontrast, learned multiparametric MR biomarker that can be used during treatment for intratreatment assessment, validated in a VX2 rabbit tumor model.
A deep convolutional neural network was trained on noncontrast multiparametric MR images using the NPV biomarker from follow-up MR imaging (3-5 days after MRgFUS treatment) as the accurate label of nonviable tissue.
A novel volume-conserving registration algorithm yielded a voxel-wise correlation between treatment and follow-up NPV, providing a rigorous validation of the biomarker. 
The learned noncontrast multiparametric MR biomarker predicted the follow-up NPV with an average DICE coefficient of 0.71, substantially outperforming the current clinical standard (DICE coefficient = 0.53). 
Noncontrast multiparametric MR imaging integrated with a deep convolutional neural network provides a more accurate prediction of MRgFUS treatment outcome than current contrast-based techniques.  
\end{abstract}

%% file: intro.tex
\section{Introduction}

Minimally and noninvasive ablation therapies for treatment of different pathologies have become promising alternatives to surgical resection. 
Ablation treatments are currently applied to several indications, including prostate disease \cite{haider2008dynamic,rouviere2012prostate,kirkham2008mr}, uterine fibroids \cite{stewart2003focused,tempany2003mr,hesley2013mr}, bone metastases \cite{gianfelice2008palliative,zaccagna2015palliative}, liver \cite{leslie2008high,wijlemans2012magnetic}, and breast \cite{sabel2004cryoablation,futureneedsHectors2016}.
Magnetic-resonance-guided focused ultrasound (MRgFUS) is a noninvasive ablation treatment. An essential component of MRgFUS procedures is accurately monitoring the treatment in real time and assessing the viability of ablated tissue immediately after the procedure \cite{futureneedsHectors2016}.  
MR provides excellent soft tissue contrast for all phases of MRgFUS procedures, including treatment planning, real-time temperature monitoring, and treatment assessment. MR biomarkers derived from MR imaging are used to assess the extent of treated tissue. If untreated tissue is detected, then additional ablation needs to be performed \cite{futureneedsHectors2016}.  

The most common MR biomarker for assessing treated tissue is the nonperfused volume (NPV) \cite{hectors2014multiparametric,futureneedsHectors2016} that is derived from contrast-enhanced (CE) T1-weighted (T1w) MR imaging.
The NPV is characterized by a lack of signal enhancement from impaired blood flow due to vascular damage and coagulation from MRgFUS ablation \cite{futureneedsHectors2016,payne2013vivo,wijlemans2013evolution,mcdannold2006uterine}.
However, the NPV biomarker obtained immediately after MRgFUS ablation treatment is not reliable for assessing the extent of the treated region due to confounding treatment effects that directly affect tissue perfusion such as edema, hyperemia, and increased vascular permeability  \cite{futureneedsHectors2016}.
Additionally, continuing treatment after CE imaging inhibits further treatment monitoring and can trap toxic contrast agent in the tissue, prohibiting further MRgFUS treatment if viable tumor tissue is detected \cite{futureneedsHectors2016,hijnen2013stability,hijnen2013magnetic}.
The need is critical for noncontrast MR biomarkers that can accurately assess tissue viability during and immediately after treatment without inhibiting further MRgFUS ablation. 

Several noncontrast, immediate MR biomarkers from multiparametric MR (MPMR) imaging have been investigated to assess MRgFUS immediately after treatment, including T2w images, MR temperature imaging (MRTI), and apparent diffusion coefficient (ADC) maps \cite{hectors2014multiparametric,plata2015feasibility,mannelli2009assessment,haider2008dynamic,wu2006registration}.
These acute MR biomarkers are based on intrinsic tissue properties that are sensitive to MRgFUS-induced changes \cite{futureneedsHectors2016}. 
However, the nonlinear relationship between MRgFUS-induced tissue property changes and tissue viability is unknown. 
Supervised machine learning provides a flexible platform to learn a nonlinear mapping between immediate MR biomarkers and tissue viability. 
Using the learned non-linear relationship would provide accurate treatment assessment during and immediately after MRgFUS treatments using noncontrast MR biomarkers. 

Supervised machine learning requires an accurate label of tissue viability. 
The NPV acquired immediately after MRgFUS ablation does not accurately identify tissue viability due to transient physiological changes.
Alternatively, the NPV biomarker 3-5 days after MRgFUS treatment (follow-up NPV) is a more accurate label of tissue viability \cite{payne2013vivo, wijlemans2013evolution, leslie2008high}.
This follow-up NPV could be used as a label for nonviable tissue to facilitate learning MPMR biomarkers for immediate MRgFUS treatment assessment. 
However, this follow-up NPV is not aligned with the immediate MPMR biomarkers due to changes in patient pose, positioning, and tissue deformation.
Image registration is necessary to account for changes in subject pose and positioning to provide a spatially accurate label of nonviable tissue for supervised machine learning.

Image registration has been an active research area for the analysis of longitudinal imaging, and numerous image registration algorithms have been proposed in the literature \cite{wu2006registration,ou2011dramms, li2009nonrigid, brock2006feasibility,ou2015deformable,jahani2018deformable}.
The objective of longitudinal registration is to correct for changes in patient position and pose between imaging sessions.
The volume of tissues such as muscle, fat, and bone is preserved under normal physiological loading, such as the changes in patient position and pose \cite{humphrey2003continuum}.
Additionally, over the first 3-5 days, minimal tissue reabsorption and healing processes after MRgFUS ablation do not result in a measurable change in tissue volume \cite{godwin2009healing}.
A volume-conserving registration must be implemented to ensure registration does not introduce artificial changes in the ablation volume in 3-5 day MR images. 
Algorithms presented in the literature regularize volume change during registration, but they do not strictly enforce volume conservation \cite{ou2015deformable,jahani2018deformable,rohlfing2003volume}.
To our knowledge, no publicly available volume conserving image registration pipeline exists. 

A longitudinal registration method that strictly enforces volume preservation is needed to allow for accurate correction of pose and positioning deformations that occur between follow-up NPV and immediate MPMR images. 
To address the specialized registration need, we developed AVOCADO: \textsc{\textbf{A} \textbf{VO}lume \textbf{C}onserving \textbf{A}lgorithm for \textbf{D}iffe\textbf{O}morphisms}, a longitudinal registration pipeline that computes a volume-preserving deformation.
AVOCADO aligns the follow-up NPV with the immediate MPMR images to yield a spatially accurate label of nonviable tissue. 

This work utilizes this novel registration pipeline to evaluate a \textsc{\textbf{M}ulti\textbf{P}arametric \textbf{B}iomarker \textbf{C}onvolutional \textbf{N}eural \textbf{N}etwork} (MPB-CNN), a deep learning model for predicting nonviable versus viable tissue from immediate MPMR imaging.
The resulting label of nonviable tissue from AVOCADO is used to train MPB-CNN to predict the nonviable tissue from immediate MPMR images without a contrast agent.
These new techniques are demonstrated using \textit{in vivo} data of a preclinical rabbit VX2 rabbit tumor model.

%% file: studydesign.tex
\begin{figure}[ht]
  \begin{center}
    \includegraphics[width=1.0\adaptivefigwidth]{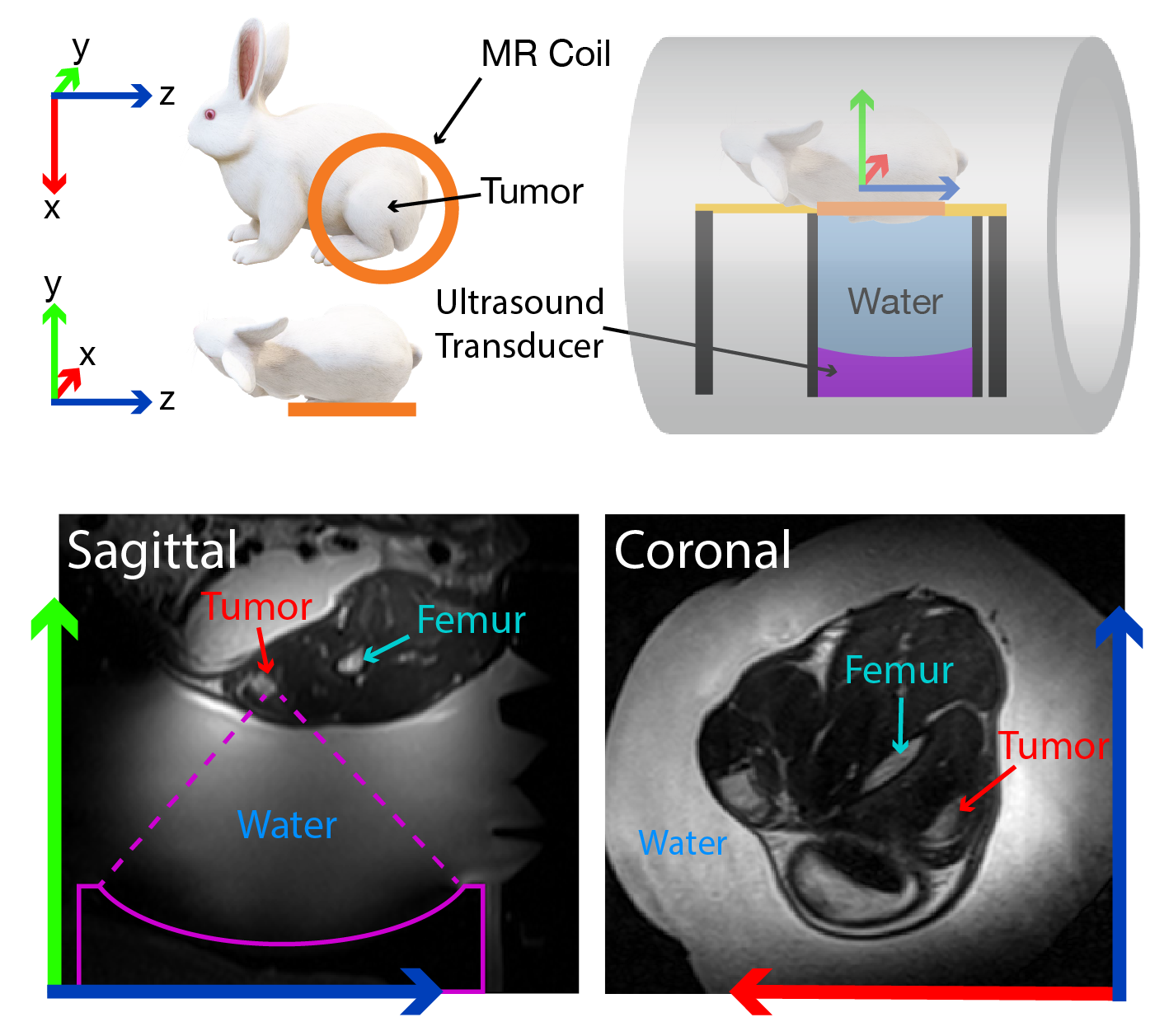}
  \end{center}
  \caption{\label{fig:setup} The top row shows the positioning of the rabbit in the MR bore and relative to the ultrasound transducer. The radiofrequency receiver MRI coil incorporated in the MR table is shown in orange relative to the animal quadriceps. The bottom row shows examples of the acquired T2-weighted images in relation to the overall setup. The colored arrows along each image indicate the MR direction. }
\end{figure} %MR setup
\setlength{\tabcolsep}{3pt}
\begin{table*}[ht]
	\centering
	\ra{1.3}
	\small
	\caption{\label{tab:params}Parameters for the multiparametric scanning sequence used for ablation day and follow-up imaging. Coronal orientation was used for all MR acquisitions. TR: Repetition Time, TE: Echo Time, EPI: Echo Planar Imaging, SPACE: Sampling Perfection with Application Optimized Contrasts Using Different Flip Angle Evolution, VIBE: Volumetric Interpolated Breath-Hold, GRE: Gradient Recalled Echo, MRTI: Magnetic Resonance Temperature Imaging, SS: Single Shot, SE: Spin Echo, and ETL: Echo Train Length. $^{*}$ T1w sequence with contrast for NPV segmentation. $^{**}$ T1w sequence without contrast for image registration. }
	\begin{tabular}{@{}l c c c c c c c c c@{}} \toprule
		\begin{tabular}
			{@{}c@{}}Scan \\ Type
		\end{tabular} & Sequence & TR (ms) & TE (ms) &\begin{tabular}
			{@{}c@{}}Flip \\ Angle
		\end{tabular} & \begin{tabular}
			{@{}c@{}}Field of \\View (mm)
		\end{tabular}  &   \begin{tabular}
			{@{}c@{}}Pixel Bandwidth \\ (Hz/Pixel)
		\end{tabular} &   \begin{tabular}
			{@{}c@{}}Acquisition \\ Resolution (mm)
		\end{tabular} &   \begin{tabular}
			{@{}c@{}}Number \\ Averages
		\end{tabular} & \begin{tabular}
			{@{}c@{}}Acquisition Time \\ (mm:ss.ms)
		\end{tabular}\\ \midrule
		MRTI & \begin{tabular} {@{}c@{}}GRE-EPI\\ (ETL=7) \end{tabular} & 25 & 11 & 14$^{\circ}$ & 192$\times$150$\times$20 & 750 & 1.5$\times$1.5$\times$2.0 &1& 0:04.50  \\
		T1w$^{*}$ & VIBE & 7.19 & 2.05 & 15$^{\circ}$ & 256$\times$192$\times$52 & 250 & 1.0$\times$1.0$\times$1.0 &1& 1:03.00  \\
		T2w & SPACE & 2000 & 300 & 120$^{\circ}$ & 256$\times$192$\times$52 & 700 & 1.0$\times$1.0$\times$1.0 &2& 5:12.00 \\
		T1w$^{**}$ & VIBE & 7.19 & 2.52 & 15$^{\circ}$ & 256$\times$192$\times$56 & 250 & 0.5$\times$0.5$\times$1.0 &3& 6:19.00  \\ 
		Diffusion & \begin{tabular} {@{}c@{}}SS-SE-EPI\\ (ETL=92)\\ (b=20,500) \end{tabular}& 7500 & 117 & 90$^{\circ}$ &160$\times$116$\times$20 & 1260 & 1.25$\times$1.25$\times$2.0 &1& 1:38.00 \\ \bottomrule
	\end{tabular}
\end{table*} %MR parameters
\section{Study Design}
\label{sec:study_design}

A rabbit tumor model \cite{palussiere2003feasibility} was used to generate a data set with and without contrast biomarkers to compare MPB-CNN against the clinical biomarker (NPV obtained immediately after treatment).
The ablation procedure and associated imaging protocol were designed to capture MPMR and structural anatomical images (noncontrast T1w) at distinct time points throughout the procedure. 
MPMR images were acquired before, during, and after MRgFUS ablation treatment to generate the immediate MPMR biomarker images. 
CE T1w MR images were collected following the MPMR imaging protocol at the end of the MRgFUS ablation procedure and used to generate the clinical biomarkers.
Following MRgFUS ablation, the animal recovered for 3-5 days post-treatment, and follow-up imaging, including the MPMR protocol and CE-T1w MR imaging, was performed to generate the follow-up NPV label of nonviable tissue. 
The specifics of the animal tumor model, ablation parameters, and image acquisitions are described below. 
The University of Utah Institutional Animal Care and Use Committee approved all procedures (\#17-08012, September 7, 2017).

\subsection{Animal Model and Hardware}
A VX2 (5x10\textsuperscript{6} cells in 50\% media/Matrigel) cell suspension was injected intramuscularly into both quadriceps of New Zealand white rabbits (N=8). 
On MRgFUS treatment day, the animal was anesthetized with a ketamine/xylazine injection (IM, 25/5 mg/kg) and intubated. 
Anesthesia was maintained with inhaled isoflurane (0.5-4.0\%).
Under anesthesia, hair was removed via clippers and a depilatory cream (Nair\textsuperscript{\texttrademark}) was applied to obtain an appropriate acoustic window for treatment. 
Using a preclinical MRgFUS system (Image Guided Therapy, Inc., Pessac, France), ablation was performed on one tumor and surrounding muscle tissue with a 256-element phased-array transducer (Imasonic, Voray-sur-l'Ognon, France; 10-cm focal length, 14.4 x 9.8 cm aperture, f = 940 kHz, 1.8x2.5x10.2 mm full width half maximum size of the intensity profile as measured in water via hydrophone (HNR-1000, Onda, Sunnyvale, CA)) inside a 3T MRI scanner (PrismaFIT Siemens, Erlangen, Germany). 
At 3-5 days post-treatment, the animal was anesthetized and follow-up MR imaging was obtained (\tableautorefname~\ref{tab:params}).
A depiction of the setup can be seen in \figureautorefname~\ref{fig:setup}.

\subsection{Ablation Parameters}
Each animal received a mean of 11 sonications (range: 8-14) during the MRgFUS ablation procedure, with a mean acoustic power of 47 W (range: 30-69 W) and an average sonication duration of 31 seconds (range: 15-40 seconds), for an average total output energy of 17,665 J (range: 11,563-26,004 J).
The number of sonications, power output, tumor volume, and total output energy are shown in \tableautorefname~\ref{tab:acoustics} for each subject. The resulting lesion sizes have large variations due to subject-specific factors including inadequate acoustic coupling with residual hair in the ultrasound path, reflection and refraction issues due to the location of the tumor with respect to bone, beam reflection due to the incidence angle of ultrasound with the skin, heterogeneous acoustic tissue properties including assumed low tumor attenuation, and animal motion.
All FUS sonications were monitored in real time with a 3D MRTI gradient echo sequence with a segmented echo planar imaging readout (\tableautorefname~\ref{tab:params}). 

\begin{table}[!b]
    \ra{1.1}
  \centering
  \caption{Ablation details for each subject including number of ablation points, acoustic power output and total energy used for each subject. }
  \small
  \begin{tabular}{ccccc}
    \toprule
    & \multicolumn{1}{c}{Tumor} & \multicolumn{1}{c}{Number of} & \multicolumn{1}{c}{Acoustic Power} &  \multicolumn{1}{c}{Total}\\
    % \cmidrule(lr){2-2}
    % \cmidrule(lr){3-4}
    % \cmidrule(lr){5-6}
  Subj. & {Vol.($mm^3$)} & {Sonications} & {Mean $\pm$ 1 Std. (W)} & {Energy (kJ)} \\
    \midrule
    1& 1477.5 &11 & 36 $\pm$18 & 16.50 \\
    2& 757.0&13 & 30 $\pm$ 8  & 11.23 \\
    3& 785.5&11 & 57 $\pm$ 17 & 23.14 \\
    4& 594.5&10 & 38 $\pm$ 5  & 12.74 \\
    5& 858.5&8 & 44 $\pm$ 7  & 11.56 \\
    6& 248.1&12 & 69 $\pm$ 18 & 26.00 \\
    7& 1929.5&14& 44 $\pm$ 9  & 18.59 \\
    8& 806.4&10 & 56 $\pm$ 9  & 18.55 \\
    \bottomrule
  \end{tabular}
  \label{tab:acoustics}
\end{table}

\subsection{Image Acquisition}
A custom, single-loop radiofrequency MR receiver coil was incorporated into the MRgFUS treatment table to improve the signal to noise ratio (SNR) of the MR imaging (\figureautorefname~\ref{fig:setup}).
MPMR images were acquired at three distinct time points: pre-ablation, immediately post-treatment, and 3-5 days post-treatment (follow-up). 
The details of each scan collected at each time point are outlined in \tableautorefname~\ref{tab:params}. 
In addition to collecting a T1w image at each time point without contrast, the same sequence was acquired immediately ($\sim$30 seconds) after administering gadolinium contrast after ablation on treatment day and at the end of follow-up imaging 3-5 days post-treatment.
The CE T1w images were collected to generate the NPV biomarker both immediately after and 3-5 days post-treatment. 
The NPVs were generated by semiautomatic segmentation of the CE images and were validated by an expert radiologist (NW).
The high-resolution T1w image without contrast acquired at all time points depicts the quadriceps anatomy without showing the tumor or ablated regions and can therefore be used for registration without biasing the registration with the tumor or ablated regions.

%% file: registration.tex
\section{Registration Methods}
\label{sec:registrationMethods}

For completeness, we describe the specialized registration that was developed to facilitate the MPB-CNN prediction of tissue viability. 
For the described data set in \sectionautorefname~\ref{sec:study_design}, the source image is the noncontrast T1w MR image acquired during follow-up imaging, and the target image is the noncontrast T1w MR image acquired immediately after the MRgFUS ablation procedure. 
The noncontrast images do not include intensity features of the tumor or ablation sites, making the intensity-based registration unbiased to these features. 
Registration estimates a high-dimensional spatially smooth one-to-one and onto transformation, or a diffeomorphism, that maps the locations from a source image to a target image, specifically: $\phi \colon \Omega \subset \mathbb{R}^3 \rightarrow \Omega$.
AVOCADO utilizes a multiscale gradient flow that is driven first by discrete landmarks and then by image intensity. 
Helmholtz-Hodge decomposition is used to guarantee volume conservation.
These sequential steps result in a volume-preserving diffeomorphism that accurately maps the follow-up NPV biomarker to the immediate MPMR images.  
This accurate voxel-wise registration enables training MPB-CNN to generate an immediate MPMR biomarker to assess the extent of MRgFUS ablation treatments without contrast imaging. 

\subsection{Gradient Flow}
A multiscale algorithm was developed for generating a diffeomorphism that maps two anatomies based on the gradient flow methodology previously used for general image registration problems \cite{christensen1996deformable,joshi2000landmark,guo2006multi,haker2004optimal, bauer2014overview}.
A diffeomorphism is constructed in terms of solutions to the Lagrangian transport equation, defined via the  ordinary differential equation (ODE) 
\begin{equation}
    \frac{d\phi(\vec x,t)}{dt} = v(\phi(\vec x,t), t), \hspace{5mm} \phi(\vec x,0) = \vec x, \hspace{5mm} t \in [0, 1]
    \label{eqn:ODE}
\end{equation}
where $\vec x$ is position, $t$ is time, and $v(x,t)$ is a time-varying velocity vector field.
The diffeomorphism $\phi(\cdot, \cdot)$ at time point $t$ is controlled by the velocity field $v(x,t)$ via the associated integral equation
\begin{equation}
    \phi(\vec x, t) = \vec x + \int_{0}^{t}v\left(\phi(\vec x,\tau),\tau\right) d\tau, \hspace{5mm} x \in \Omega.
    \label{eqn:integral}
\end{equation}
If the velocity field is sufficiently well behaved~\cite{younes2010shapes,misiolek1993stability}, $\phi(\vec x,t)$ is guaranteed to be a diffeomorphism. 
The inverse diffeomorphism $\phi^{-1}(\vec x,t)$ satisfies the Eulerian ODE given by
\begin{equation}
    \frac{d\phi^{-1}(\vec x,t)}{dt} = - D\left[\phi^{-1}(\vec x,t)\right]v(\phi^{-1}(\vec x,t), t)
\end{equation}
where $D\left[ \cdot \right]$ is the Jacobian.  
The above ODE can be approximately integrated via an implicit Euler method:
\begin{equation}
    \phi^{-1}(\vec x, t+ \delta_t) = \phi^{-1}(x - \delta_t v(\vec x, t), t)
    \label{eqn:update}
\end{equation}
where $\delta_t$ is a scalar time step.

Using the transport equation above, a gradient flow is developed, governed by forcing functions $g(x, t)$.
The forcing functions are the gradient of an energy potential that defines the registration problem.
A forcing function $g(x,t)$ is related to the velocity field via
\begin{equation}
    \label{eqn:forcing}
    Lv(\phi(\vec x, t)) = g(\phi(\vec x, t))
\end{equation}
where $L$ is a partial differential operator that has an associated smoothing kernel $K$ such that
\begin{equation}
    v(\phi(\vec x, t)) = Kg(\phi(\vec x, t)).
\end{equation}
The operator $K$ ensures the fields $v(x,t)$ are smooth and gives sufficient differentiability for the existence and uniqueness of the solution to the ODE, which guarantees that $\phi(\vec x, t)$ is a diffeomorphism  \cite{joshi2000landmark}. 
Properties of the final diffeomorphism can be controlled via the properties of $v$ at every time step.  

\subsection{Helmholtz-Hodge Decomposition}
\label{sec:HHD}
Under normal physiological tissue loading, soft tissues exhibit incompressible behavior due to high water volume fractions \cite{humphrey2003continuum}.
Enforcing that the velocity fields $v(x, t)$ in \equationautorefname~(\ref{eqn:integral}) describe incompressible fluid flow ensures that the  diffeomorphism is volume preserving. 
Mathematically, incompressibility is equivalent to ensuring the divergence of $v(x, t)$ is zero: $\nabla \cdot v(x, t) = 0$ \cite{hinkle20124d}. 
According to the Helmholtz-Hodge decomposition, a smooth and rapidly decaying vector field can be uniquely decomposed into an orthogonal sum of an irrotational, or curl-free, vector field and a solenoidal, or divergence-free, vector field \cite{bhatia2012helmholtz}. 
Uniquely decomposing the velocity field $v(x, t)$ allows the calculation of the solenoidal component by removing the irrotational component, which projects $v(x, t)$ into the space of divergent-free vector fields \cite{cantarella2002vector}.
By enforcing that each velocity field governing the ODE in \equationautorefname~(\ref{eqn:ODE}) is divergent free with the Helmholtz-Hodge decomposition, the deformation is constrained to be volume preserving and models realistic soft tissue deformation.

The Helmholtz-Hodge decomposition and vector field projection of a discretized velocity field is implemented in the Fourier domain. 
A vector field is projected to the space of solenoidal vector fields by removing the curl-free component: 
\begin{equation}
    \vec F_{curl \; v}(\vec \omega) = \vec F_v({\vec \omega}) - \left( \frac{\vec W(\vec \omega) \cdot \vec F_v(\vec \omega)}{\norm{\vec W(\vec \omega)}}\right) \frac{\vec W(\vec \omega)}{\norm{\vec W(\vec \omega)}}.
\end{equation}
where $\vec W(\cdot)$ is the Discrete Fourier Transform (DFT) of the discrete gradient operator and $\vec \omega$ is the angular frequency.
The inverse DFT of $F_{curl \; v}(\vec \omega)$ gives a divergent-free velocity field $v(x, t)$ for solving \equationautorefname~(\ref{eqn:ODE}).
Controlling the divergence of the velocity field at every iteration generates a final diffeomorphism with physiologically expected incompressibility.
Additionally, the incompressibility can be spatially modulated using a spatially varying convex combination of the original velocity field and the projected velocity field as follows:
\begin{equation}
    \label{eqn:spatial_vary}
    v(x, t) = (1 - \alpha(x)) \; v_{orig}(x, t) \: + \: \alpha(x) \; v_{divfree}(x, t)
\end{equation}
where $v(x, t)$ is the final velocity field, $v_{orig}(x, t)$ is the original unconstrained velocity field and $v_{divfree}(x, t)$ is the projected divergence free velocity field.
When $\alpha(x) = 1$, the original volume is conserved point wise over the entire image domain. 
This study utilized $\alpha(x) = 1$.

\begin{figure}[t]
\centering
\begin{minipage}{\adaptivefigwidth}
    \centering\hspace{-2mm}
    \raisebox{-.55\height}{%
    \begin{subfigure}[b]{0.34\columnwidth}
        \centering
        \begin{tikzpicture}
        \draw (0, 0) node[inner sep=0] {\includegraphics[
        width=\columnwidth,
        trim={1in 1.3in 1in 1.05in},clip
        ]{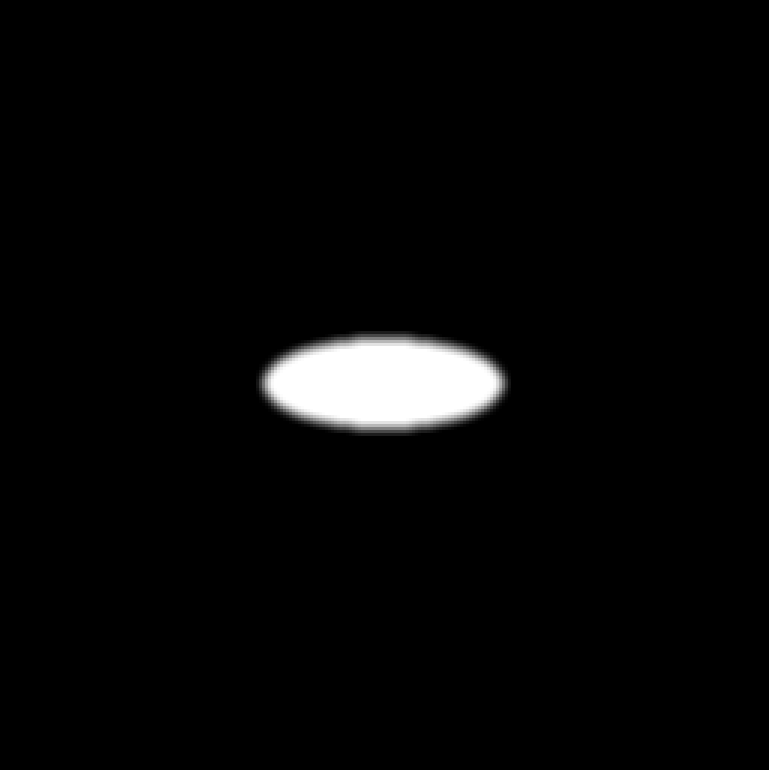}};
        \draw (0.0,0.7) node[text=white]{\small \textbf{A = 1888.00 px}};
        \end{tikzpicture}
        \subcaption*{Source Image}
    \end{subfigure}}%
    \hspace{0.4cm}
    \huge$\xleftrightharpoons[\phi^{-1}]{\phi}$
    \hspace{0.0cm}
    \raisebox{-.55\height}{%
    \begin{subfigure}[b]{0.34\columnwidth}
        \centering
        \begin{tikzpicture}
        \draw (0, 0) node[inner sep=0] {\includegraphics[
        width=\columnwidth,
        trim={1in 1.3in 1in 1.05in},clip
        ]{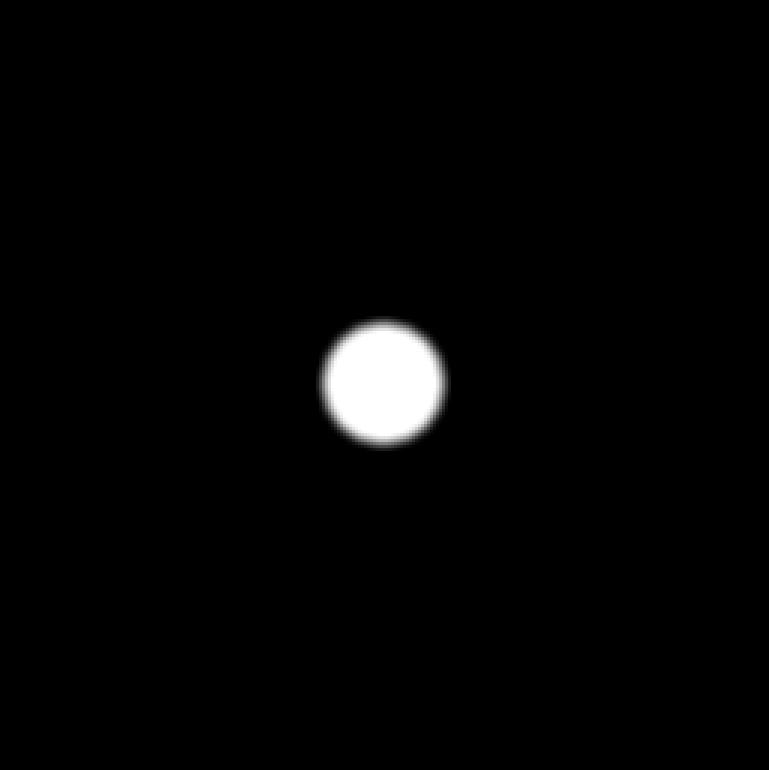}};
        \draw (0.0,0.7) node[text=white]{\small \textbf{A = 1264.00 px}};
        \end{tikzpicture}
        \subcaption*{Target Image}
    \end{subfigure}}
    \\\vspace{1mm}\hspace{-5mm}%
    \begin{subfigure}[c]{0.285\columnwidth}
        \begin{tikzpicture}
            \draw (0, 0) node[inner sep=0] {\includegraphics[
            width=\columnwidth,
            trim={1in 1.3in 1in 1.05in},clip
            ]{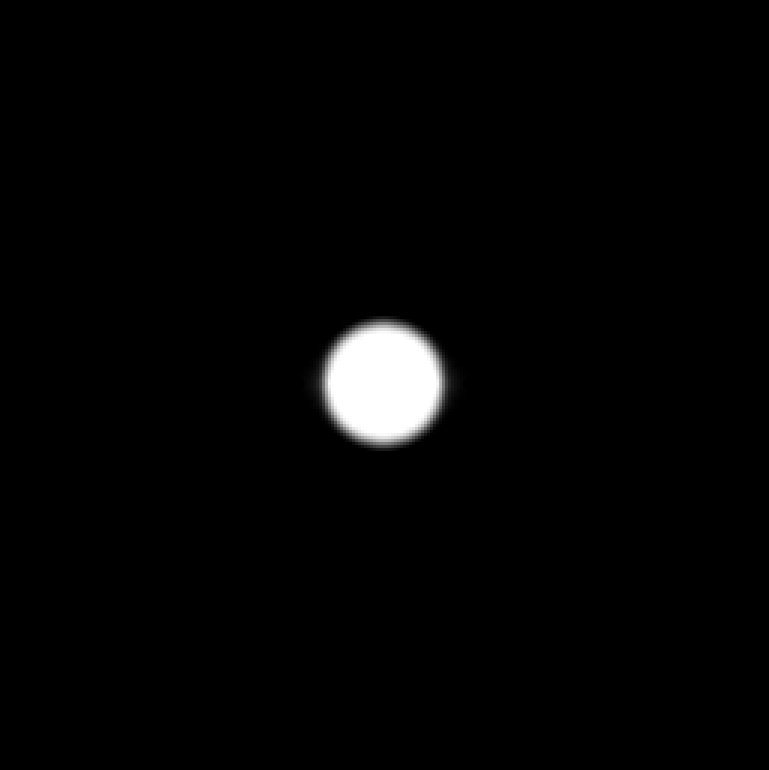}};
            \draw (0.0,0.7) node[text=white,inner sep=0]{{\small\textbf{A = 1267.44 px}}};
        \end{tikzpicture}
    \end{subfigure}
    \begin{subfigure}[c]{0.285\columnwidth}
        \begin{tikzpicture}
            \draw (0, 0) node[inner sep=0] {\includegraphics[
        width=\columnwidth,
        trim={1in 1.3in 1in 1.05in},clip
        ]{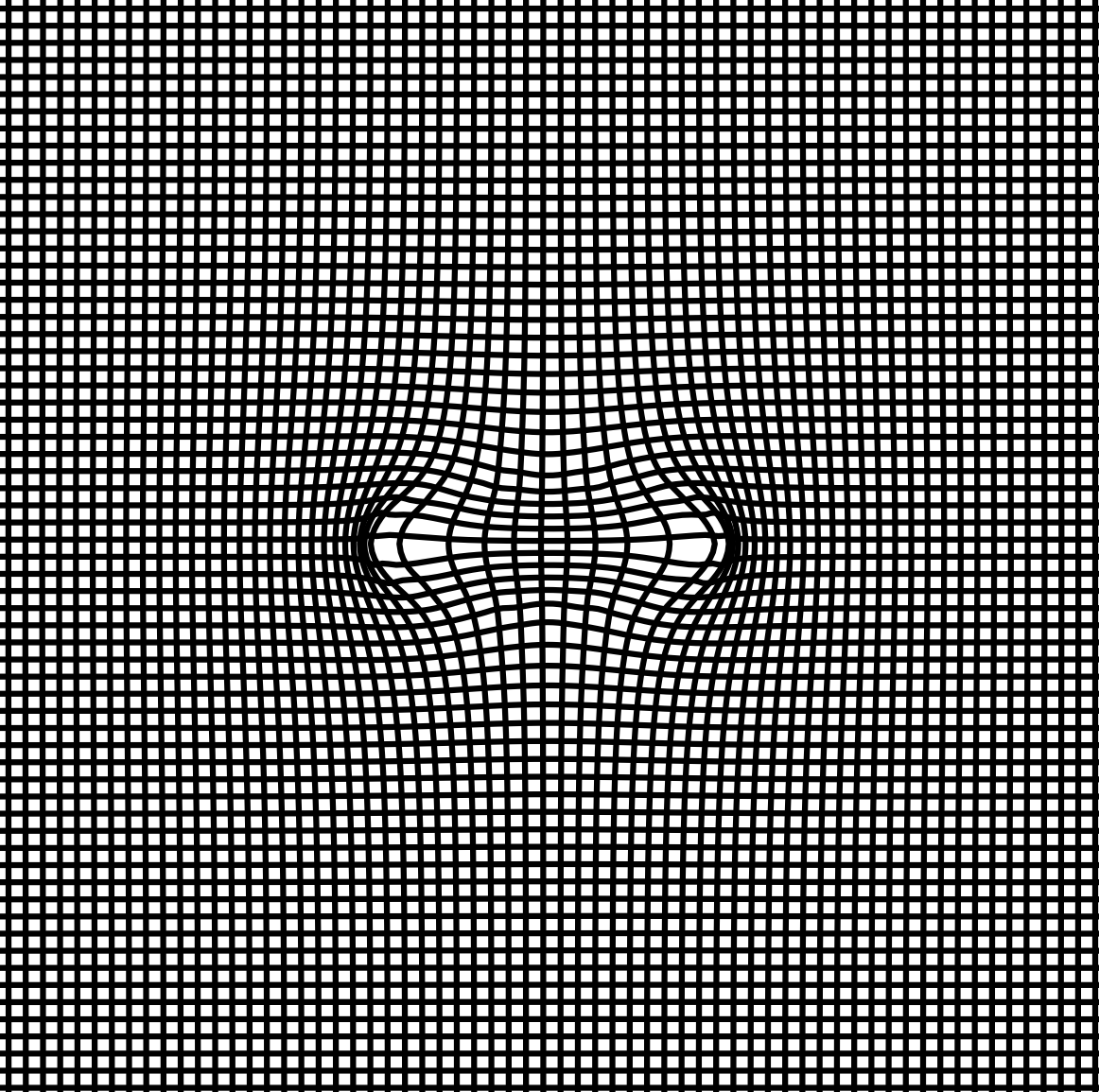}};
        \end{tikzpicture}
    \end{subfigure}
    \begin{subfigure}[c]{0.285\columnwidth}
        \frame{\includegraphics[
        width=\columnwidth,
        trim={1in 1.3in 1in 1.05in},clip
        ]{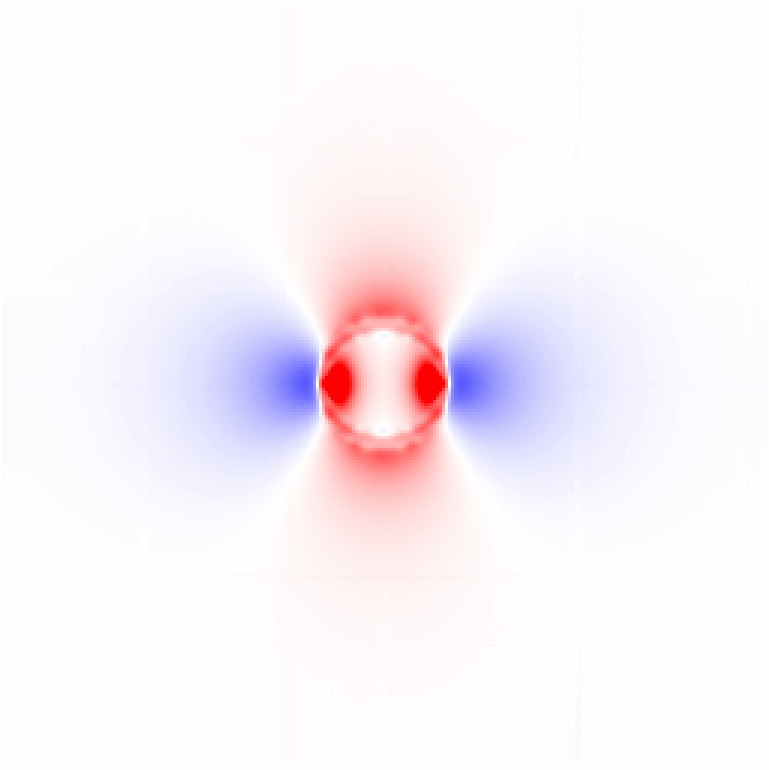}}
    \end{subfigure}%
    \begin{subfigure}[c]{0.02\columnwidth}
        \begin{tikzpicture}
            \draw (0, 0) node[inner sep=0] {\includegraphics[
            width=0.6\columnwidth,
            height=0.80in,
            trim={2.0in 0in 0in 0in},clip
            ]{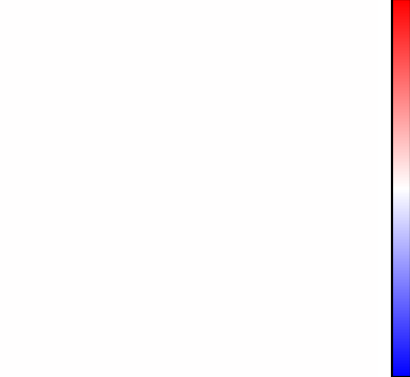}};
            \draw (0.35,0.9) node[text=black]{\small 2.0};
            \draw (0.35,0.0) node[text=black]{\small 1.0};
            \draw (0.36,-0.9) node[text=black]{\small 0.0};
        \end{tikzpicture}
    \end{subfigure}
    \\\vspace{-2mm}\hspace{-5mm}%
    \begin{subfigure}[c]{0.285\columnwidth}
        \begin{tikzpicture}
            \draw (0, 0) node[inner sep=0] {\includegraphics[
            width=\columnwidth,
            trim={1in 1.3in 1in 1.05in},clip
            ]{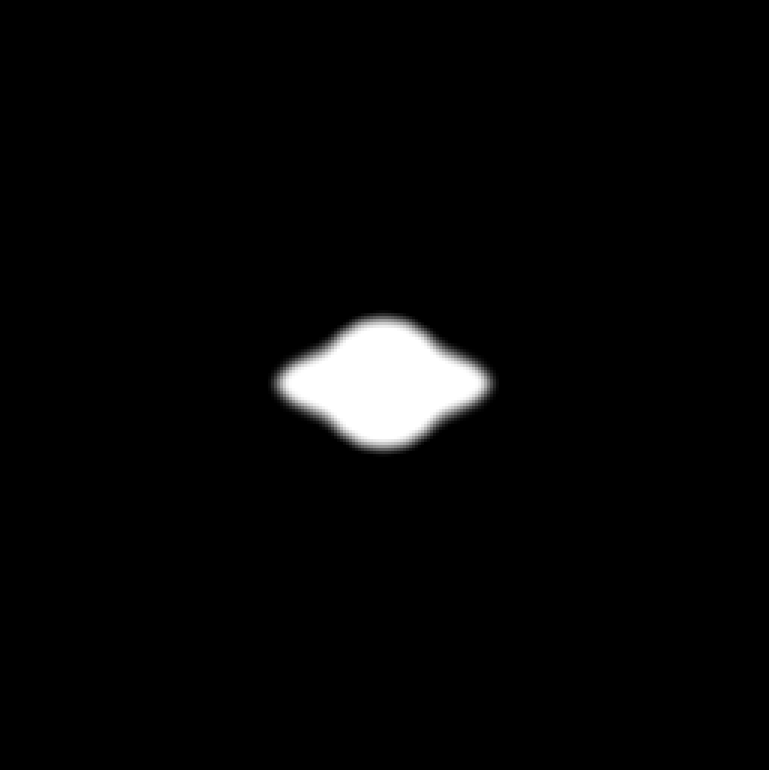}};
            \draw (0.0,0.7) node[text=white,inner sep=0]{{\small\textbf{A = 1887.36 px}}};
        \end{tikzpicture}
        \caption*{$I(\phi^{-1})$ Image}
    \end{subfigure}
     \begin{subfigure}[c]{0.285\columnwidth}
        \begin{tikzpicture}
            \draw (0, 0) node[inner sep=0] {\includegraphics[
        width=\columnwidth,
        trim={1in 1.3in 1in 1.05in},clip
        ]{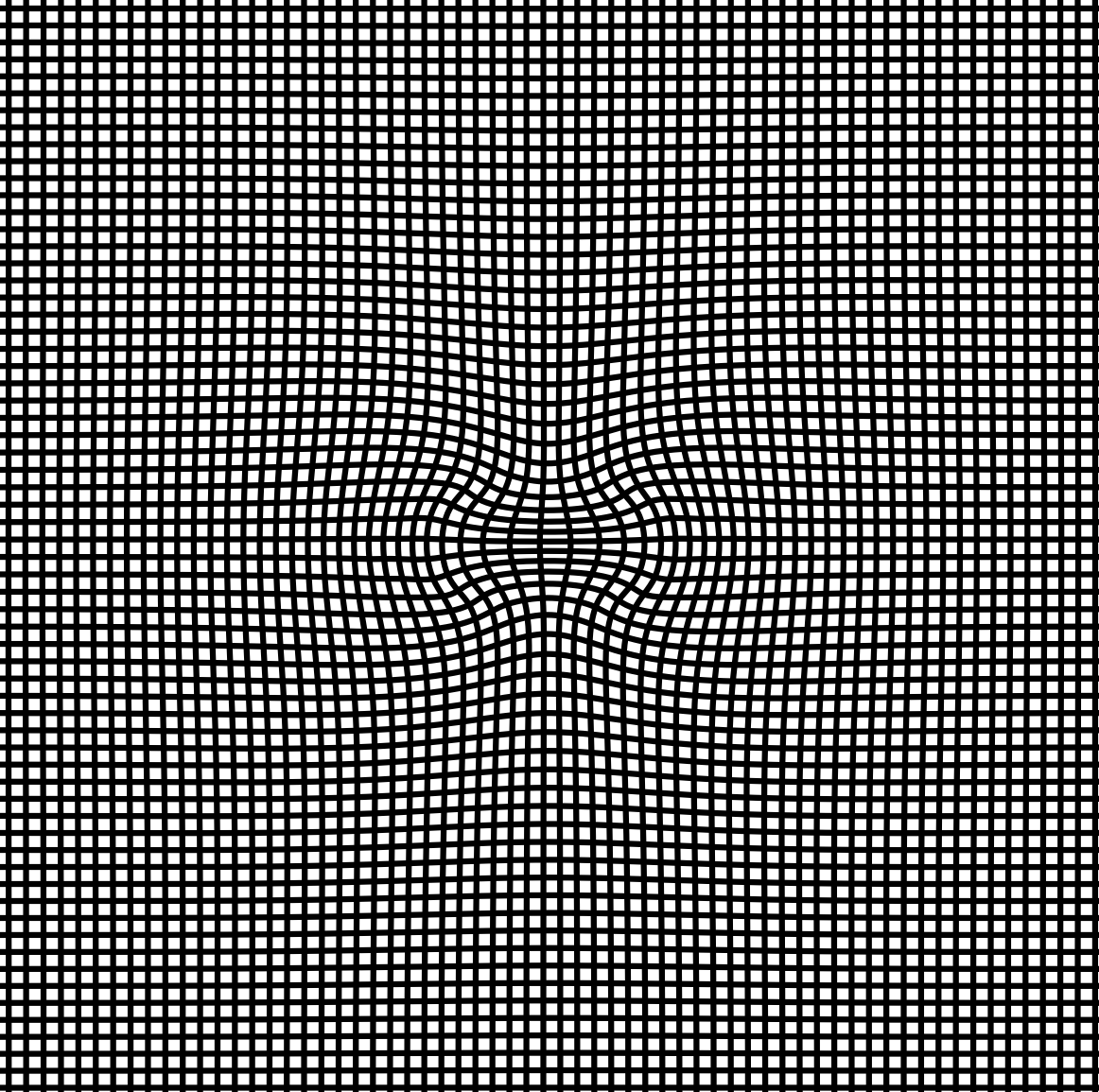}};
        \end{tikzpicture}
        \caption*{$\phi^{-1}$ Field}
    \end{subfigure}
    \begin{subfigure}[c]{0.285\columnwidth}
        \frame{\includegraphics[
        width=\columnwidth,
        trim={1in 1.3in 1in 1.05in},clip
        ]{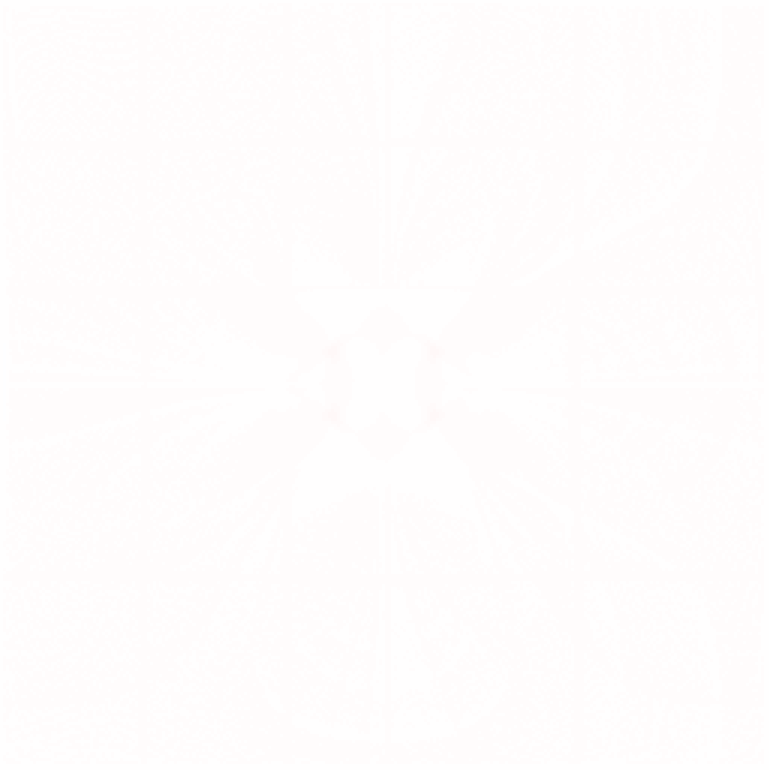}}
        \caption*{$|D\phi^{-1}|$}
    \end{subfigure}%
    \begin{subfigure}[c]{0.02\columnwidth}
        \begin{tikzpicture}
            \draw (0, 0) node[inner sep=0] {\includegraphics[
            width=0.6\columnwidth,
            height=0.80in,
            trim={2.0in 0in 0in 0in},clip
            ]{./Figures/Incompressible/colorbar.png}};
            \draw (0.35,0.9) node[text=black]{\small 2.0};
            \draw (0.35,0.0) node[text=black]{\small 1.0};
            \draw (0.36,-0.9) node[text=black]{\small 0.0};
        \end{tikzpicture}
        \caption*{}
    \end{subfigure}
    \end{minipage}%
    \caption{Example of non-volume-preserving versus volume-preserving gradient flow registration. The top row shows the source and target image with their respective areas (A) in pixels. The middle and bottom rows from left to right show the deformed image, deformation field, and Jacobian determinant for non-volume-preserving (middle row) and volume-preserving registration (bottom row), respectively. } 
    \label{fig:incomp_comp}
\end{figure}

An example of registration both with and without incompressibility can be seen in \figurename~\ref{fig:incomp_comp}. 
The top row shows the source and target images and their respective starting areas (A) in units of pixels (px).
The middle row shows the result of registration \textit{without} projecting the velocity field $v(x,t)$ to the space of divergent-free velocity fields.
The area of the source object is reduced by 32.87\% to match the target area.
However, if the ellipse is a biological tissue, then the area should remain constant through registration because tissues are incompressible. 
The bottom row shows the volume-preserving result of registration with projecting the velocity field $v(x, t)$ to the space of divergent-free velocity fields. 
The volume-preserving method generates deformations that reflect the incompressibility of biological tissues.

 % with and without volume pres

\subsection{Rigid Registration}
\label{sec:rigid}
In the outlined study design, exactly replicating subject pose between treatment and follow-up imaging was impossible.
For large changes in the subject pose, AVOCADO registration begins with a landmark registration problem based on user-defined source and target anatomical landmarks selected in follow-up and treatment-day coordinates, respectively.
Anatomical landmarks are used to estimate a volume-preserving rigid transformation in two steps. 
First, an affine transformation $A$ is estimated via a least squares minimization:
\begin{equation}
    A = \left(\vec c_i^{} \vec p_i^{\;T}\right)\left(\vec p_i^{} \vec p_i^{\;T}\right)^{-1}
\end{equation}
where $\vec p_i$ are the source landmarks and $\vec c_i$ are the target landmarks. 
Second, the affine transformation is projected to a rigid transformation $R$ via the singular value decomposition of the matrix $A$ \cite{van1983matrix}. 

\subsection{Landmark-Driven Gradient Flow}
\label{sec:landmarks}
Given the rigid transformation $R$, an anatomical landmark-driven higher dimensional transformation is estimated by using gradient flow on an energy potential $g(x) = E_{RBF}$, from \equationautorefname~(\ref{eqn:forcing}), which is a function only of the landmark points:
\begin{equation}
    E_{RBF} = \sum_{i=1}^{M} \norm{R \vec p_i - \phi^{-1}(\vec c_i, t)}^2
\end{equation}
where $M$ is the number of landmark pairs and $\phi^{-1}(\cdot, t)$ is the inverse diffeomorphism at time $t$. 
A multiquadratic radial basis function (RBF) is used as the smoothing operator $K$:
\begin{equation}
	 K(\vec r) = \sqrt{1 + \left(\epsilon \; \vec r\right)^2}, \hspace{5mm} \vec r = \vec x - \vec c
\end{equation}
where $\epsilon$ is a shape-tuning parameter, set to $\epsilon=1.0$ for all experiments in this work. 
Although thin plate splines are commonly used, they are ill-conditioned \mbox{without} an affine transformation.
However, an affine transformation is not volume preserving. 
The velocity field $v(x, t)$ is expressed as a summation of weighted RBF fields.
Given the distance $d$ between points $\vec d_i = Rp_i - \phi^{-1}(\vec c_i, t)$, the weights $b$ are solved via
\begin{equation}
	\begin{bmatrix}
	b_1(t) \\
	\vdots \vspace{1mm}\\
	b_i(t)
	\end{bmatrix}
 =
\begin{bmatrix}
	K(\vec r_{11})  & \dots    & K(\vec r_{1j})\\
    \vdots                & \ddots   & \vdots\\
	K(\vec r_{i1})  & \dots    & K(\vec r_{ij})\\ 
\end{bmatrix} ^{-1}
	\begin{bmatrix}
	\vec d_1(t) \\
	\vdots \vspace{1mm}\\
	\vec d_i(t)
	\end{bmatrix}
	\label{eqn:RBF_weights}
\end{equation}
with $\vec r_{ij} = \phi^{-1}(\vec c_i, t) - \phi^{-1}(\vec c_j, t)$ \cite{buhmann2003radial}. 
The velocity field is given by
\begin{equation}
    v(\vec x, t) = \sum_{i=1}^{M} b_i(t) K\left(\norm{R \vec x - \vec c_i}\right)
    \label{eqn:RBF}
\end{equation}

The vector field $v(\vec x, t)$ is subsequently projected into the space of divergent-free vectors fields via the Helmholtz-Hodge decomposition described in \sectionautorefname~\ref{sec:HHD}.
For a given time step, the diffeomorphism $\phi^{-1}$ is updated via
\begin{equation}
    \phi^{-1}(\vec x, t+\delta_t) = \phi^{-1}(x - \delta_t v(\vec x, t), t)
    \label{eqn:rbf_update}
\end{equation}
where $\delta_t$ is a scalar step size. 
For a given time step, we update the diffeomorphism according to \equationautorefname~(\ref{eqn:update}) until a convergence criterion is met.
For matching landmarks, the convergence point is when the $E_{RBF}$ becomes less than the average interuser variability ($\epsilon_u$) of selecting the landmarks, which is determined from the test-retest capability of users when selecting the source anatomical landmarks (\sectionautorefname~\ref{sec:landmark_val}). 

\subsection{Image Intensity-Driven Gradient Flow}
\label{sec:images}
The final registration step is an image registration problem to match two anatomies based on image similarities of a source and target image. 
The image intensity-driven gradient flow is initialized with the diffeomorphism from landmark-driven gradient flow. 
The energy potential $E_{Image}$ for the image registration problem is
\begin{equation}
    E_{Image} = \int_{\Omega} \norm{I_1(\phi^{-1}(\vec x, t)) - I_0(\vec x)}^2
    \label{eqn:image_energy}
\end{equation}
where $I_1$ is the source image (follow-up, noncontrast T1w VIBE image) and $I_0$ is the target image (treatment-day, noncontrast T1w VIBE image). 
The associated forcing function from \equationautorefname~(\ref{eqn:forcing}) is given by
\begin{equation}
    g(\vec x, t) = \left( I_1(\phi^{-1}(\vec x, t)) - I_0(\vec x) \right) \nabla I_1(\phi^{-1}(\vec x, t))
    \label{eqn:energy}
\end{equation}
where $\nabla$ is the gradient operator. 

An operator of the Cauchy-Navier type is chosen for the $L$ operator from \equationautorefname~(\ref{eqn:forcing}) with $L= -\alpha \Delta + \gamma I$, where $\Delta$ is the Laplace operator and $I$ is the identity operator. 
The scalars $\alpha$ and $\gamma$ control smoothness and ensure the operator is nonsingular, respectively. 
This operator has been used previously in image registration \cite{beg2005computing}.
The smoothing kernel associated with $L$ is applied in the Fourier domain with $\text{DFT}\{K\} = \text{DFT}\{L\}^{-1}$ to get $v(x, t)$.
The smoothed vector field is projected into the space of divergent-free vector fields using the Helmholtz-Hodge decomposition described in \sectionautorefname~\ref{sec:HHD}. 
For a given time step, the diffeomorphism is updated according to \equationautorefname~(\ref{eqn:update}).
The intensity-based registration continues until the change in energy $E_{Image}$ in \equationautorefname~(\ref{eqn:image_energy}) from the previous iteration is less than $\epsilon_I = 3 \times 10^{-4}$.

In total, the final diffeomorphism defined by AVOCADO is the composition of the rigid transformation, the gradient flow on anatomical landmarks, and the gradient flow on the image intensities.
An outline of the final algorithm is shown in \figureautorefname~\ref{alg:algo}.
This volume-preserving method ensures that registration process does not bias the volume of the follow-up NPV biomarker.
\begin{figure}[tb]
\centering
\begin{minipage}{0.48\textwidth}
	\centering
	\small
	\begin{algorithmic}[1]
	    \setstretch{1.25}
		\Procedure{AVOCADO}{$ \: p_i, c_i, I_0, I_1, \delta_t \: $}
		\State Solve affine $A = \left(c_i^{}p_i^T\right)\left(p_i^{}p_i^T\right)^{-1}$
		\State Singular Value Decomposition $A \mapsto R$
		\State $\phi^{-1}_{RBF}(\vec x, k=0) := \vec x$
		\While{$E_{RBF} \geq \epsilon_u, k=k+1$}
		    \State Solve spline weights $b = K^{-1}d$
		    \State $v(\vec x, k) = \sum_{i=1}^{M} b_i K\left(\norm{R \vec x - \phi^{-1}(\vec c_i, k)}\right)$
		    \State Project div-free $\delta_t v(\vec x, k) \mapsto s(\vec x, k)$
		    \State $\phi^{-1}_{RBF}(\vec x, k + 1) = \phi^{-1}_{RBF}(\vec x - s(\vec x, k), k)$ 
		\EndWhile
		\State $\phi^{-1}_{I}(\vec x, k = 0) = \phi^{-1}_{RBF}(\vec x, k)$
		\While{$|E_{Image}(k - 1) - E_{Image}(k)| \leq \epsilon_I$}
		    \State $g(\vec x, k) = \left( I_1(\phi^{-1}_{I}(\vec x,k)) - I_0 \right) \nabla I_1(\phi^{-1}_{I}(\vec x, k))$
		    \State $v(\vec x, k) = DFT^{-1}\!\left\{ DFT\{L\}^{-1}DFT\{g(\vec x, k)\}\right\}$
		    \State Project div-free $\delta_t v(\vec x, k) \mapsto s(\vec x, k)$
		    \State $\phi^{-1}_{I}(\vec x, k + 1) = \phi^{-1}_{I}(\vec x - s(\vec x, k), k)$ 
		\EndWhile
		\State \textbf{return} $\phi_I^{-1}(\vec x, k)$
		\EndProcedure
	\end{algorithmic}
    \end{minipage}
	\caption{AVOCADO algorithm to calculate the diffeomorphism $\phi^{-1}(\vec x)$ mapping two anatomies $I_1$ and $I_0$ while preserving volume.}
	\label{alg:algo}\vspace{-0.3cm}
\end{figure}

%% file: validation.tex
\section{Registration Validation}
To ensure the accuracy and volume preservation of our registration method, we describe the methods to assess different aspects of the AVOCADO algorithm. 
The accuracy and volume preservation of our method will be compared with a state-of-the-art registration method.  

\subsection{Landmark-Based Registration Accuracy}
\label{sec:landmark_val}

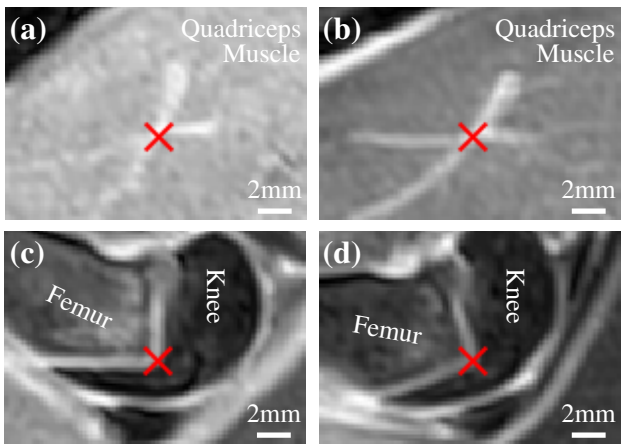
\begin{figure}[b!]
\centering
\begin{minipage}{\adaptivefigwidth}
    \centering
    \begin{subfigure}[b]{0.49\columnwidth}
        \begin{tikzpicture}
        \draw (0, 0) node[inner sep=0] {\includegraphics[
        width=\columnwidth,
        trim={0in 0.35in 0in 0.1in},clip
        ]{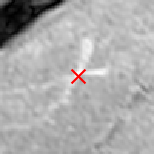}};
        \draw (-1.7,1.1) node[text=white]{\large \textbf{(a)}};
        \draw[white, line width = 2pt] (1.35,-1.3) -- (1.8,-1.3);
        \draw ((1.575,-1.0) node[text=white]{2mm};
        \draw ((1.13,1.1) node[text=white]{Quadriceps};
        \draw ((1.4,0.8) node[text=white]{Muscle};
        \end{tikzpicture}
    \end{subfigure}
    \hfill
    \begin{subfigure}[b]{0.49\columnwidth}
        \begin{tikzpicture}
        \draw (0, 0) node[inner sep=0] {\includegraphics[
        width=\columnwidth,
        trim={0in 0.35in 0in 0.1in},clip
        ]{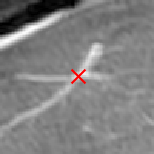}};
        \draw ((-1.7,1.1) node[text=white]{\large \textbf{(b)}};
        \draw[white, line width = 2pt] (1.35,-1.3) -- (1.8,-1.3);
        \draw ((1.575,-1.0) node[text=white]{2mm};
        \draw ((1.13,1.1) node[text=white]{Quadriceps};
        \draw ((1.4,0.8) node[text=white]{Muscle};
        \end{tikzpicture}
    \end{subfigure}
    \\\vspace{1.5mm}
    \begin{subfigure}[b]{0.49\columnwidth}
        \begin{tikzpicture}
        \draw (0, 0) node[inner sep=0] {\includegraphics[
        width=\columnwidth,
        trim={0in 0.35in 0in 0.1in},clip
        ]{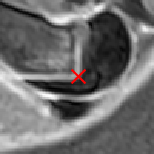}};
        \draw ((-1.7,1.1) node[text=white]{\large \textbf{(c)}};
        \draw ((-1.0,0.4) node[text=white,rotate=-25]{Femur};
        \draw ((0.8,0.55) node[text=white,rotate=-90]{Knee};
        \draw[white, line width = 2pt] (1.35,-1.3) -- (1.8,-1.3);
        \draw ((1.575,-1.0) node[text=white]{2mm};
        \end{tikzpicture}
    \end{subfigure}
    \hfill
    \begin{subfigure}[b]{0.49\columnwidth}
        \begin{tikzpicture}
        \draw (0, 0) node[inner sep=0] {\includegraphics[
        width=\columnwidth,
        trim={0in 0.35in 0in 0.1in},clip
        ]{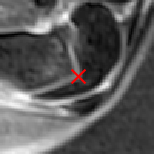}};
        \draw ((-1.7,1.1) node[text=white]{\large \textbf{(d)}};
        \draw ((-1.1,0.1) node[text=white,rotate=-10]{Femur};
        \draw ((0.6,0.55) node[text=white,rotate=-90]{Knee};
        \draw[white, line width = 2pt] (1.35,-1.3) -- (1.8,-1.3);
        \draw ((1.575,-1.0) node[text=white]{2mm};
        \end{tikzpicture}
    \end{subfigure}
    \end{minipage}
    \caption{Example validation landmarks: (a) and (b) show a corresponding blood vessel bifurcation in the treatment and follow-up CE images, respectively, that was not visible during registration; and (c) and (d) show a corresponding bone structure in the treatment and follow-up images, respectively, that was visible during registration.}
    \label{fig:landmarks}
\end{figure}

The overall accuracy of AVOCADO will be evaluated using manually selected anatomical landmarks.
Multiple observers (N=3) were given anatomical landmarks in the target images and asked to find the corresponding landmarks in the source images using 3D Slicer software \cite{fedorov20123d}.
Ten source and target landmark pairs were chosen for each subject over the entire region of interest. 
Examples of the validation landmarks can be seen in \figureautorefname~\ref{fig:landmarks}. 
The TRE was calculated by deforming anatomical landmarks with the estimated deformation and computing the Euclidean distance between the deformed point and its corresponding target landmark.
The landmarks for validation were chosen independently from the RBF initialization landmarks.

The landmark-based accuracy of AVOCADO was compared with another leading, general purpose registration algorithm to demonstrate the overall accuracy of AVOCADO.
Deformable Registration via Attribute Matching and Mutual-Saliency Weighting (DRAMMS) is the leading registration method for longitudinal registration. 
DRAMMS has been shown to consistently outperform other registration methods.  
In previous publications, DRAMMS has outperformed 12 competing registration methods, including ANTS and Demons, in overall target registration error and region similarity measures \cite{ou2011dramms, ou2014comparative, ou2015deformable, ou2012validation}.
To test the relative accuracy, DRAMMS was executed on the same noncontrast registration images used during the AVOCADO pipeline with default input parameters. 
The resulting volume change and the TRE were computed to compare DRAMMS with AVOCADO.
A two-sample-related t-test on the Euclidean error between deformed and target landmarks was used to test for a significant difference between DRAMMS and AVOCADO.

The error due to user variability when manually selecting anatomical landmarks was evaluated. 
Multiple observers selected validation landmarks, and each observer was asked to repeat their landmark selection. 
The inter- and intra-observer variations were calculated by comparing the Euclidean distance between selected landmarks both across observers and within a single observer, respectively.
A two-sample-related t-test between inter- and intra-observer errors was used to test for significance difference between the same observer and across observers.

\subsection{Changes in User-Dependent Inputs}

AVOCADO is dependent on the input anatomical landmarks for rigid and RBF transformations.
Due to manual user selection, variance is expected in the initialization landmarks introduced by the user's ability to select the anatomical landmarks. 
The robustness of AVOCADO to changes in the input landmarks was tested by applying a range of zero mean, normally distributed perturbations to each source landmark individually and executing the registration with the updated landmarks.
Changes less than the re-test capability of the observers (intraobserver error) should not change the final registration accuracy. 

\subsection{Volume Preservation Capability}

Transformations produced from AVOCADO should preserve volume when applied to follow-up MR imaging.
The volume preservation capability of our model was validated using expert segmentations of the NPV biomarker on follow-up MR imaging \cite{rohlfing2003volume}.
The volume was calculated by integrating the binary segmentation and scaling by the product of the voxel spacing. 
The estimated deformation was then applied to the follow-up NPV segmented images to register them with the MR images obtained immediately after MRgFUS ablation.
The segmentation volume before and after deformation was calculated and compared to determine the final volume change for each subject.  
A volume change of less than 0.5\% was assumed to adequately preserve volume.%

\input{gridcompfig}

%% file: gridcompfig.tex
\begin{figure*}[!b]
\vspace*{-14pt}
\centering
\small
\newlength{\mylength}
\setlength{\mylength}{0.21\textwidth}
\newlength{\myheight}
\setlength{\myheight}{0.8in}
\newcolumntype{R}{>{\centering\arraybackslash}m{0.2cm}}
\newcolumntype{C}{>{\centering\arraybackslash}m{1.01\mylength}} 
\newcolumntype{L}{>{\centering\arraybackslash}m{0.08cm}}
\begin{tabular}{RC!{\vrule width 1.5pt}RCCCL}
    \rotatebox{90}{Target} &
     \frame{\scalebox{1}[-1]{\includegraphics[
        width=\mylength,
        height=\myheight,
        trim={2.5in 0.3in 1.8in 0.1in},clip]        {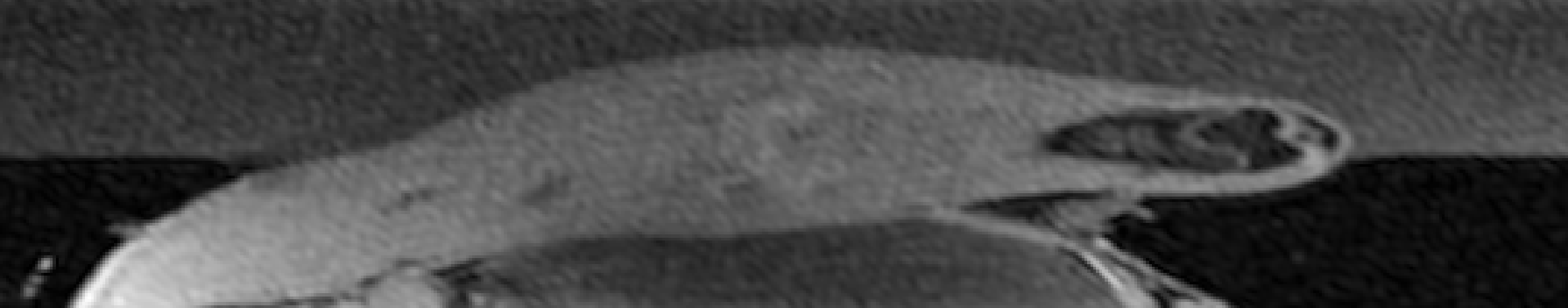}}} & 
    \rotatebox{90}{DRAMMS} &
     \frame{\scalebox{1}[-1]{\includegraphics[
        width=\mylength,
        height=\myheight,
        trim={2.0in 0.3in 1.4in 0.1in},clip]
        {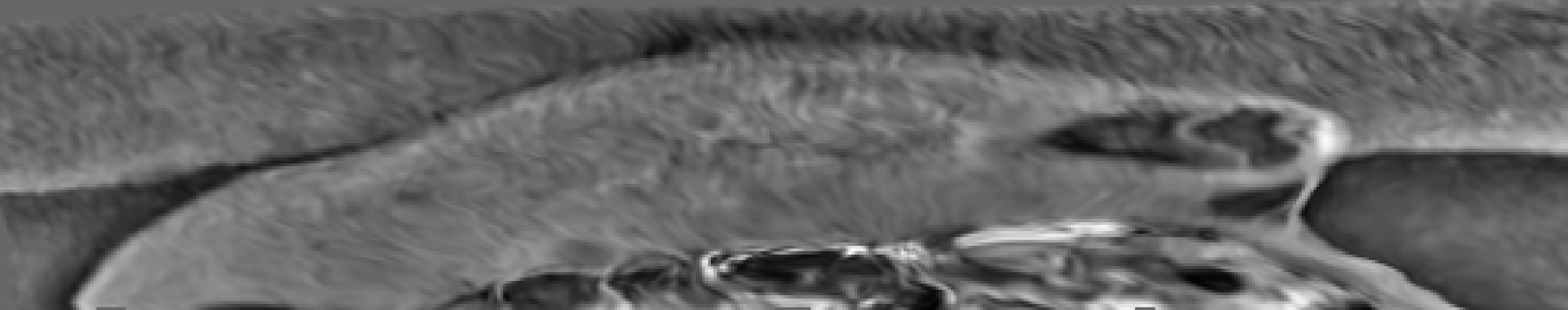}}} & 
     \frame{\scalebox{1}[-1]{\includegraphics[
        width=\mylength,
        height=\myheight,
        trim={2.0in 0.3in 1.4in 0.1in},clip]
        {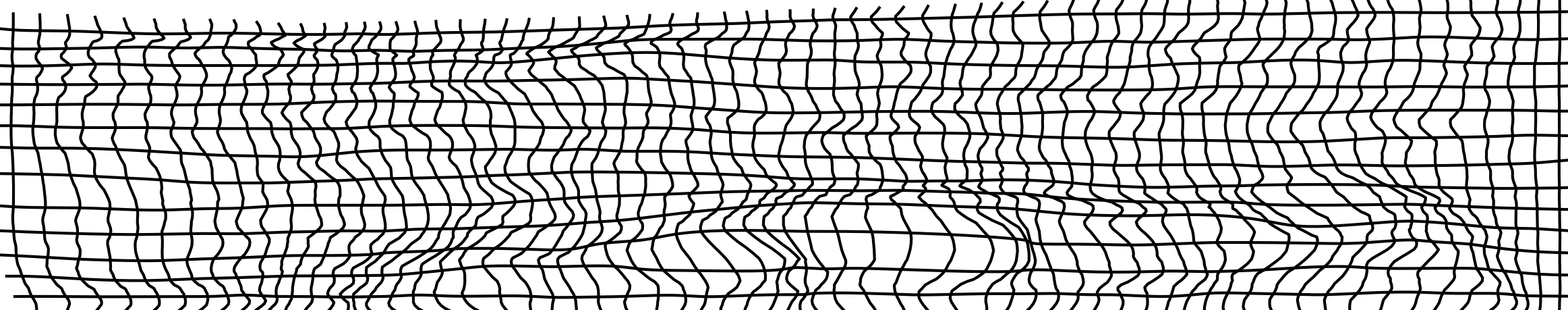}}} & 
    \frame{ \scalebox{1}[-1]{\includegraphics[
        width=\mylength,
        height=\myheight,
        trim={2.0in 0.3in 1.4in 0.1in},clip]
        {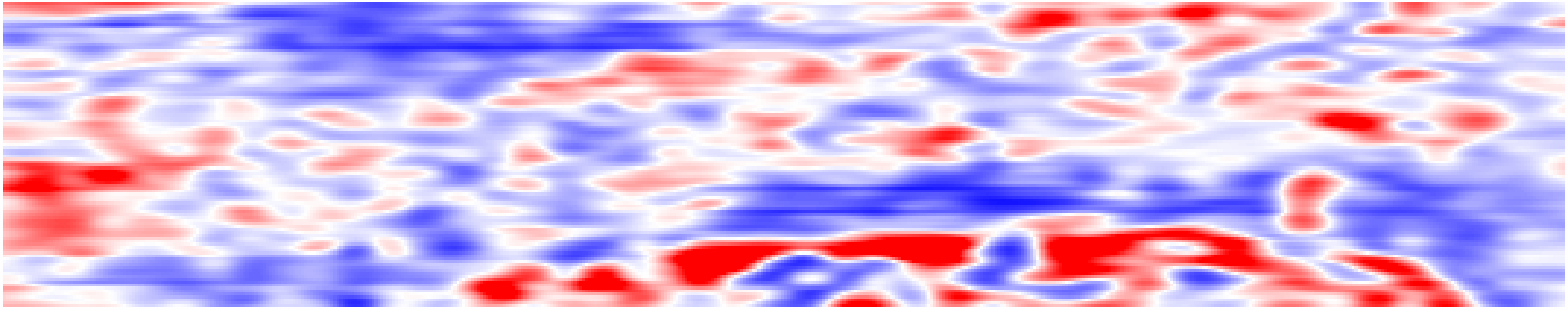}}} &
    \begin{tikzpicture}
        \draw (0.0, 0) node[inner sep=0] {\includegraphics[
        width=0.02\mylength,
        height=\myheight,
        trim={2.0in 0in 0in 0in},clip
        ]{./Figures/Incompressible/colorbar.png}};
        \draw (0.28,0.92) node[text=black]{\footnotesize 2.0};
        \draw (0.28,0.0) node[text=black]{\footnotesize 1.0};
        \draw (0.28,-0.92) node[text=black]{\footnotesize 0.0};
    \end{tikzpicture}\\
    \vspace{-4pt}
     & Noncontrast Images & & Deformed Images & $\phi^{-1}$ Fields & Jacobian Determinants & \\
     \rotatebox{90}{Source} &
     \frame{\scalebox{1}[-1]{\includegraphics[
        width=\mylength,
        height=\myheight,
        trim={2.0in 0.3in 1.4in 0.1in},clip]
        {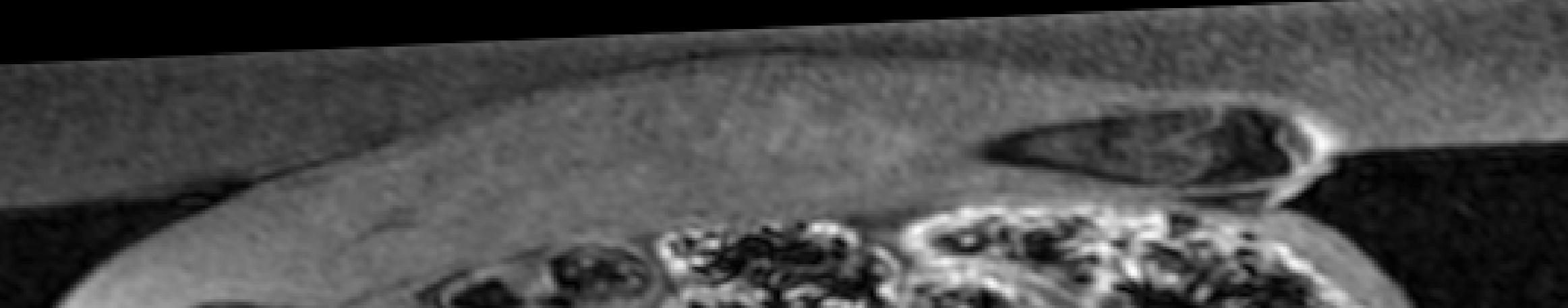}}} &
     \rotatebox{90}{AVOCADO} &
     \frame{\scalebox{1}[-1]{\includegraphics[
        width=\mylength,
        height=\myheight,
        trim={2.0in 0.3in 1.4in 0.1in},clip]
        {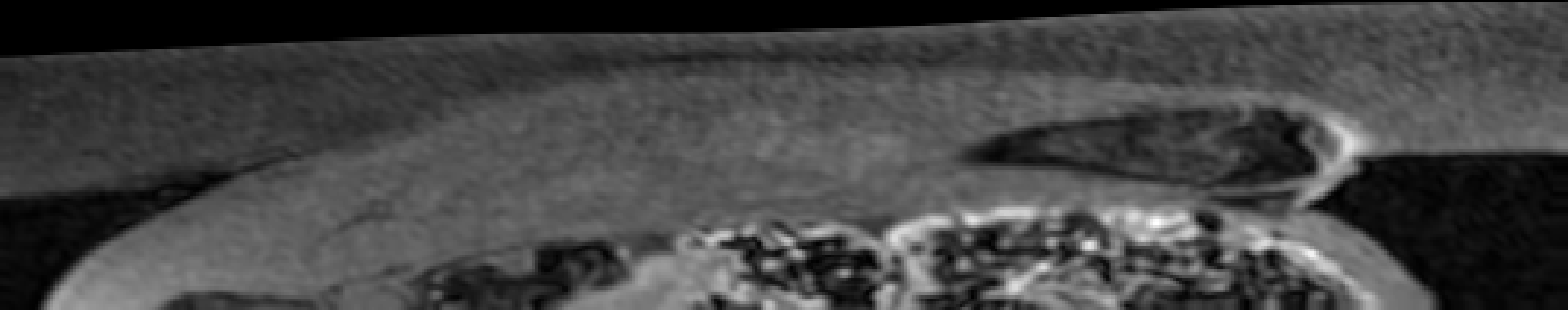}}} & 
     \frame{\scalebox{1}[-1]{\includegraphics[
        width=\mylength,
        height=\myheight,
        trim={2.0in 0.3in 1.4in 0.1in},clip]
        {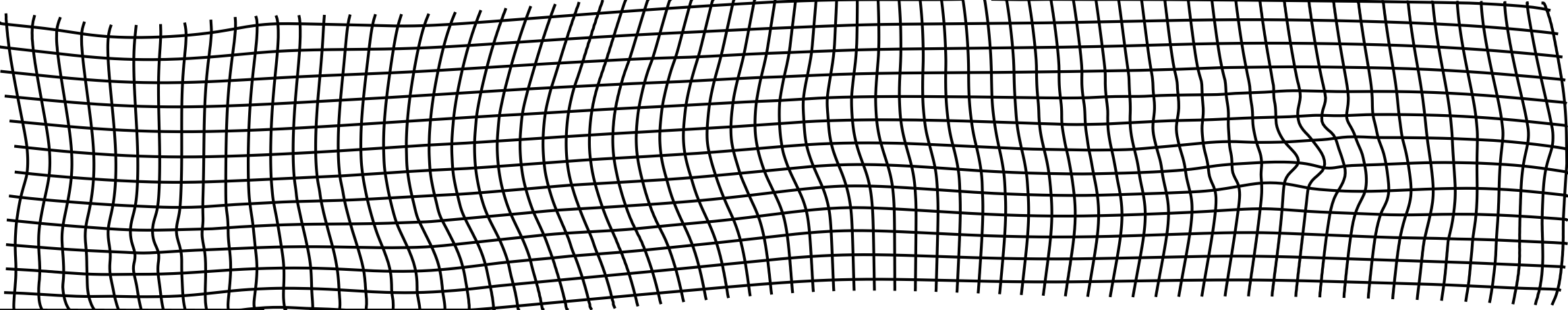}}} & 
     \frame{\scalebox{1}[-1]{\includegraphics[
        width=\mylength,
        height=\myheight,
        trim={2.0in 0.3in 1.4in 0.1in},clip]
        {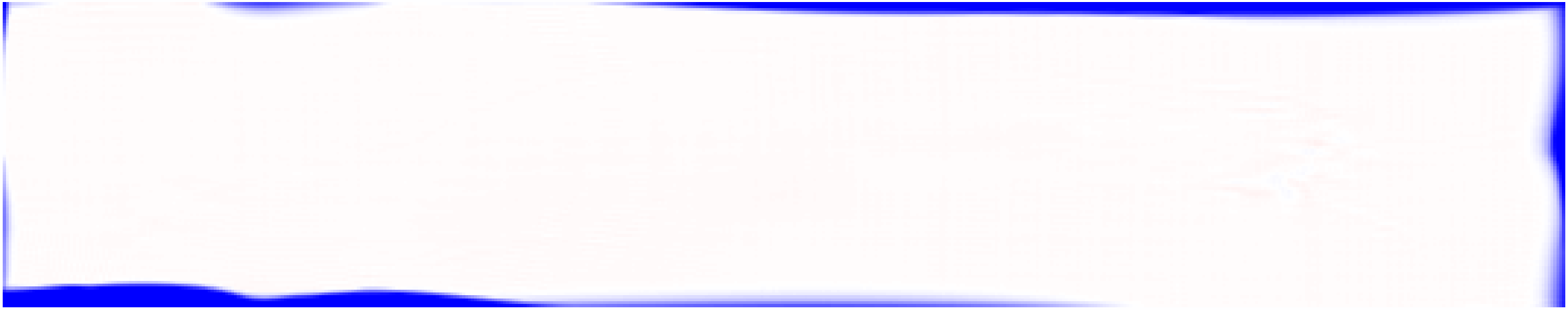}}} & 
    \begin{tikzpicture}
            \draw (0.0, 0) node[inner sep=0] {\includegraphics[
            width=0.02\mylength,
            height=\myheight,
            trim={2.0in 0in 0in 0in},clip
            ]{./Figures/Incompressible/colorbar.png}};
            \draw (0.28,0.92) node[text=black]{\footnotesize 2.0};
            \draw (0.28,0.0) node[text=black]{\footnotesize 1.0};
            \draw (0.28,-0.92) node[text=black]{\footnotesize 0.0};
        \end{tikzpicture}
\end{tabular}
\vspace*{-4pt}
\caption{Subject-specific (subject 7) comparison of DRAMMS and AVOCADO deformation fields. The first column shows the noncontrast source and target images used for registration. The second column shows the deformation field produced by DRAMMS (top) and AVOCADO (bottom) after registering the source and target images. Note the smoothness of the deformation and homogeneity of the Jacobian determinant produced by AVOCADO.%for the volume-preserving registration.
}
\label{fig:grid_comp}
\end{figure*} % Grid comparison

%% file: network.tex
\section{MultiParametric Biomarker CNN}

The input data for the MPB-CNN are the pre- and post-MRgFUS ablation MPMR images described in \tableautorefname~\ref{tab:params}.
The label was the follow-up NPV registered with the treatment-day MPMR images according the AVOCADO algorithm. 
The objective was to train the MPB-CNN to predict the nonviable tissue using only noncontrast MPMR images acquired pre- and post-MRgFUS ablation. 
The performance of the MPB-CNN was evaluated by comparing the prediction with the treatment-day NPV, which is a commonly used clinical metric of nonviable tissue. 
The network was considered successful if its prediction of the nonviable tissue was as accurate or more accurate than the treatment-day NPV as measured by the DICE coefficient \cite{dice1945measures}. 
The data set, preprocessing, and neural network architecture for deep learning MR biomarkers are described here. 

\subsection{Data Set and Preprocessing}
The training data for MPB-CNN was selected from the results of AVOCADO registration between immediate MPMR and follow-up images.
Two subjects were excluded from the MPB-CNN set due to small follow-up NPV volumes, leaving N=6 animals for MPB-CNN data. 
In the six subjects used for training and validation, there was no notable growth in the tumor size in the 3-5 day period following treatment.
Each acute MR image was resampled with linear interpolation onto a common grid with 0.5 mm isotropic resolution to accommodate the subset of immediate MPMR images acquired at lower resolutions. 
The target label for each subject was the expert segmentation of the follow-up NPV deformed with the subject-specific diffeomorphism from AVOCADO, which results in a common grid for all the real-time MR images and target images for each subject. 

The six MR images input to the network were 2D slices from: (1) pre- and (2) postablation T2w images, (3) pre- and (4) postablation apparent diffusion coefficient (ADC) maps, (5) cumulative thermal dose (CTD) map, and (6) a maximum temperature projection (MTP) map, derived from MR temperature imaging. 
Although not inclusive of all MPMR contrasts, these inputs were chosen due to the resolution achievable in reasonable time limits (5-10 minutes per contrast) and prior work showing changes in these parameters to be indicative of ablation damage \cite{hectors2014multiparametric}.
The ADC maps were calculated from diffusion weighted imaging using the MRI vendor's proprietary image reconstruction pipeline (Siemens, Erlangen, Germany).
The CTD maps were calculated from MRTI using the equivalent number of minutes of heating at 43 $\degree$C relationship commonly used for thermal therapies \cite{van2013cem43, dewhirst2003basic}. 
The MTP maps were the highest temperature recorded for each pixel from the MRTI over time (using a rectal probe for starting temperature).

All images were cropped to a $128 \times 128 \times 90$ region centered on the registered follow-up NPV segmentation, and image intensities were normalized using the average maximum from that image type across all subjects (e.g., the average maximum ADC map value across all subjects is used to normalize all the ADC maps, etc.). 
The center slice along the last dimension of each $128 \times 128 \times 90$ volume was extracted for evaluation.
However, the slices adjacent to the center slice are too similar to the evaluation slice to include in the training data. 
The four slices on either side of the evaluation slice were omitted from the testing and training data in an attempt to ensure the network is not biased. 
As a result, we had 486 ($[90 - 9] \times 6$ animals) $128 \times 128$ 2D samples for training and 6 $128 \times 128$ 2D samples for evaluation.
Each input training batch was augmented with: 1) 0-10\% random changes in the brightness, contrast, and saturation; 2) random horizontal and vertical flips; 3) random rotations of 0-20 degrees; and 4) zero mean Gaussian noise. 

\subsection{Network Architecture}
A U-Net segmentation architecture was used for the MPB-CNN with six input channels \cite{ronneberger2015u}.
In addition to the superior performance of U-Net compared to other CNN architectures, this architecture has demonstrated success with a small number of training samples, making it ideal for the MPB-CNN. 
The network returned a pixel-wise probability of the tissue viability, and the output of the network was evaluated against the follow-up NPV segmentation using a binary cross entropy loss function. 
The network was validated with the central slice from each input volume slab (which was excluded from the training data). 
When evaluating the network, the probability output from the network was thresholded at a value of 0.5 to yield a segmentation of viable versus nonviable tissue. 
The purpose of the MPB-CNN was to provide a more accurate and immediate prediction of treated tissue than the treatment-day NPV biomarker without using CE imaging.
Comparison of the MPB-CNN biomarker with the treatment day NPV biomarker was computed by the DICE coefficient between each MPB-CNN prediction against the follow-up NPV segmentation.

%% file: results.tex
\section{Results}
The overall goal of registration was to align the follow-up NPV with the acute MR imaging obtained immediately after MRgFUS ablation while preserving volume.
\figureautorefname~\ref{fig:grid_comp} shows a 2D slice of the 3D deformation grid from both DRAMMS and AVOCADO and the corresponding relative volume change (Jacobian determinant) of each field. 
Both methods were executed on the same noncontrast source and target images.
The left column shows the target (top) and source (bottom) noncontrast images, the middle column shows the deformation grid, and the right column shows the Jacobian determinant, which indicates volume change relative to the starting volume. 
Note that AVOCADO produces a smoother deformation grid and preserves the original volume (Jacobian determinant of 1.0). 

\subsection{Registration Accuracy}

The final TRE for DRAMMS and AVOCADO is plotted in \figureautorefname~\ref{fig:TRE}.
For the eight rabbits analyzed in this study, the final TRE for AVOCADO ranged from 1.08 mm to 1.60 mm (mean $\pm$ STD = 1.33 $\pm$ 0.16 mm). 
For DRAMMS, the final TRE ranged from 1.07 mm to 2.78 mm (1.69 $\pm$ 0.64 mm).
An example of the deformations produced from DRAMMS and AVOCADO can be seen in \figureautorefname~\ref{fig:grid_comp}. 
The deformations $\phi^{-1}$ from AVOCADO are smoother than from DRAMMS, and the Jacobian determinant for AVOCADO is 1.0 over the entire volume.
The TRE for AVOCADO was significantly lower than DRAMMS ($p = 0.018$) while preserving the original volume of the anatomies. 

The intra-observer variability  for the validation landmarks over the eight rabbits ranged from 0.82 mm to 1.22 mm (0.93 $\pm$ 0.13 mm).
The inter-observer variability ranged from 0.65 mm to 1.26 mm (0.89 $\pm$ 0.18 mm).
There was no significant difference between the inter-observer and intra-observer errors ($p = 0.95$).
\definecolor{cat1_color}{RGB}{214,111,172}%
\definecolor{cat2_color}{RGB}{0,148,42}%
\definecolor{cat4_color}{RGB}{0,147,194}%
\definecolor{cat3_color}{RGB}{255,199,61}%
\begin{figure}[tb]
\centering
\begin{minipage}{\adaptivefigwidth}
    \centering
\begin{tikzpicture}
\tikzstyle{every node}=[font=\small]
\begin{axis}[
    width=0.9\columnwidth,
    scale only axis=true,
    height=4cm,
    xtick={1,...,8},
    xticklabels={%
        1,
        2,
        3,
        4,
        5,
        6,
        7,
        8},
    ymajorgrids,
    major tick length = 0,
    ylabel = Final TRE (mm),
    xlabel={Subject Number},
    ybar,
    legend style={anchor=north east, legend cell align=left, align=left, draw=black}%, legend columns=2
    ]
\addplot[
    fill=cat1_color,
    draw=black,
    point meta=y,
    every node near coord/.style={inner ysep=5pt},
    error bars/.cd,
        y dir=both,
        y explicit
] 
table [x=num, y error=ms, y=mm, col sep=comma] {./Figures/Error_figure/stats.csv};
\addlegendentry{AVOCADO}
\addplot[
    fill=cat2_color,
    draw=black,
    point meta=y,
    every node near coord/.style={inner ysep=5pt},
    error bars/.cd,
        y dir=both,
        y explicit
] 
table [x=num, y error=ds, y=dm, col sep=comma] {./Figures/Error_figure/stats.csv};
\addlegendentry{DRAMMS}
\draw ({rel axis cs:0,0}|-{axis cs:0,0}) -- ({rel axis cs:1,0}|-{axis cs:0,0});
\end{axis}
\end{tikzpicture}
\end{minipage}
    \caption{Plot of the TRE for both non-volume-preserving (DRAMMS) and volume-preserving (AVOCADO) registration methods.}
    \label{fig:TRE}
\end{figure}
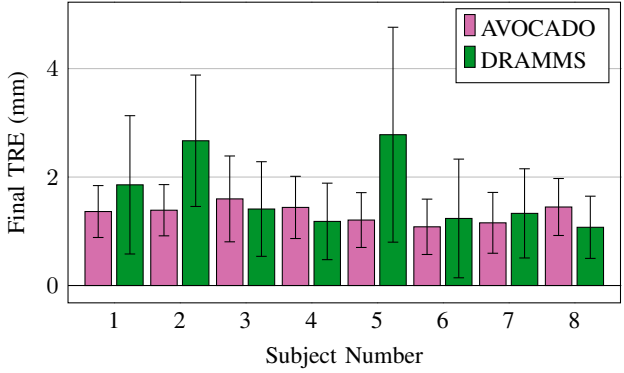 % TRE bar plot

\subsection{User-Dependent Model Inputs}

\begin{figure}[b]
\centering
\begin{minipage}{\adaptivefigwidth}
\centering
\begin{tikzpicture}
    \tikzstyle{every node}=[font=\small]
    \begin{axis}[
      xmin=0, xmax=5,
      ymin=0, ymax=5,
      ymajorgrids=true,
      grid style=dashed,
      legend pos=north west,
      ytick={0, 2,...,16},
      xtick={0, 2,...,16},
      width=\columnwidth,
      xlabel={Initialization Landmark Mean Perturbation (mm)},
      ylabel={Final Mean TRE (mm)},
      ytick style={draw=none},
      height=2.5in
    ]
    \addplot [color=purple, line width=0.5mm] table [x=xmean, y=ymean, col sep=comma]{./Figures/Robust/stats_0.1.csv};
    \addplot [name path=p1_top,color=purple!50, opacity=0.2, forget plot] table [x=xmean, y=ypstd, col sep=comma]{./Figures/Robust/stats_0.1.csv};
    \addplot [name path=p1_bot,color=purple!50, opacity=0.2, forget plot] table [x=xmean, y=ymstd, col sep=comma]{./Figures/Robust/stats_0.1.csv};
    \addplot[purple!50,fill opacity=0.2, forget plot] fill between[of=p1_top and p1_bot];
    \draw [dashed, line width=0.5mm] (0.93,0.0) -- (0.93,2.3);
    \node[draw, align=center] at (1.0,3.0) {Mean User\\Repeatability\\0.93 mm};
    \end{axis}
\end{tikzpicture}
\end{minipage}
\caption{Mean anatomical perturbation across all subjects plotted versus the final mean TRE of the registration. The mean intraobserver variability for selecting anatomical landmarks is 0.93 mm.}
\label{fig:robust}\vspace{-0.4cm}
\end{figure}
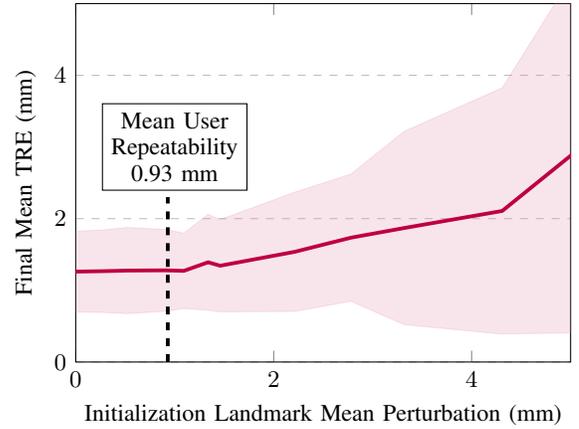

\figureautorefname~\ref{fig:robust} shows the results of perturbing the RBF initialization landmarks. 
For perturbations under the average user-repeatability (intraobserver) error, the final TRE changed minimally. 
As expected, if the perturbations became large (> 2 mm), the final TRE also increased. 
% Initialization figure

\subsection{Volume Preservation}
The original volumes, deformed volumes, and the percent change for both the non-volume-preserving (DRAMMS) and volume-preserving (AVOCADO) registration methods are reported in \tableautorefname~\ref{tab:volume_change}.
The original volume of the expert NPV biomarker segmented on follow-up CE T1w MR imaging ranged from 0.02 ml to 3.96 ml (1.34 $\pm$ 1.39 ml). 
The absolute volume change after registration with AVOCADO ranged from 0.07\% to 0.39\% (0.28 $\pm$ 0.11\%). 
All volume changes for AVOCADO are under our 0.5\% threshold for discretization or interpolation error.
\setlength{\tabcolsep}{4pt}
\begin{table}[!b]
  \ra{1.1}
  \centering
  \caption{Volumes and changes from non-volume-preserving (DRAMMS) and volume-preserving (AVOCADO) registration. A negative percent means a reduction in the final volume.}
  \small
  \begin{tabular}{cSSSSS}
    \toprule
    & \multicolumn{1}{c}{Original} & \multicolumn{2}{c}{DRAMMS} & \multicolumn{2}{c}{AVOCADO} \\
    \cmidrule(lr){2-2}
    \cmidrule(lr){3-4}
    \cmidrule(lr){5-6}
  Subj. & {Vol $mm^3$} & {Vol $mm^3$} & {\%} & {Vol $mm^3$} & {\%} \\
    \midrule
    \vspace{-3mm}
    \begin{filecontents*}{./Figures/NPV_registration/changes.csv}
num, orig, comp, compp, incomp, incompp
1,1334.75,905.31,-32.17,1339.66,0.37
2,38.00,55.77,46.76,38.13,0.34
3,915.38,903.44,-1.30,911.82,-0.39
4,161.00,145.09,-9.88,161.57,0.36
5,20.38,12.61,-38.11,20.42,0.22
6,1949.15,1614.37,-17.18,1955.99,0.35
7,2628.12,3268.78,24.38,2629.99,0.07
8,3959.38,3878.74,-2.04,3966.28,0.17
\end{filecontents*}
\csvreader[head to column names]{./Figures/NPV_registration/changes.csv}{}
    {\\\num & \orig & \comp & \compp & \incomp & \incompp}\\
    \bottomrule
  \end{tabular}
  \label{tab:volume_change}
\end{table} % Volume table

In every subject, registration with DRAMMS caused a larger change in the follow-up NPV segmentation volume when compared to the AVACADO results.
The average absolute volume change of the NPV biomarker due to registration with DRAMMS ranged from 1.30\%  to 46.76\%  (21.48 $\pm$ 16.80 \%). 
DRAMMS registration resulted in both increased or decreased NPV biomarker volume, with the largest increase of 46.47\% and the largest decrease of 38.11\%.

\subsection{MPB-CNN}

\begin{figure}[tb]
\centering
\begin{minipage}{\adaptivefigwidth}
\centering
\small
\newlength{\mynlength}
\setlength{\mynlength}{0.40\columnwidth}
\newlength{\mynheight}
\setlength{\mynheight}{0.65in}
\begin{tabular}{cc}
        \begin{tikzpicture}
        \draw (0, 0) node[inner sep=0] {\includegraphics[
        width=\mynlength,
        height=\mynheight,
        trim={0.4in 1.0in 0.6in 1.0in},clip]
        {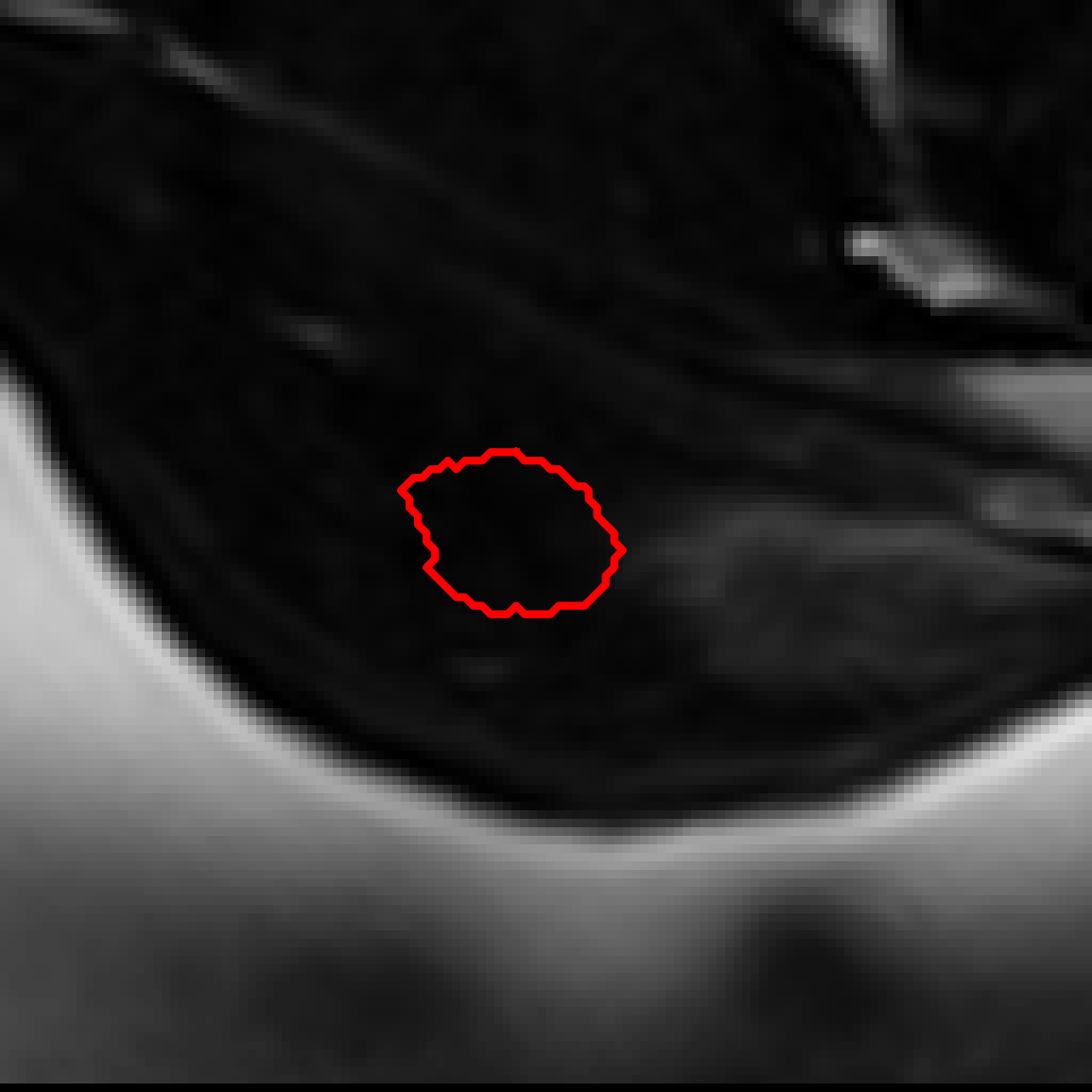}};
        \draw (-1.3,0.5) node[text=white]{\large \textbf{(a)}};
        \draw[white, line width = 2pt] (0.85,-0.6) -- (1.5,-0.6);
        \draw ((1.15,-0.4) node[text=white]{6mm};
        \end{tikzpicture} & 
     \begin{tikzpicture}
        \draw (0, 0) node[inner sep=0] {\includegraphics[
        width=\mynlength,
        height=\mynheight,
        trim={0.4in 1.0in 0.6in 1.0in},clip]
        {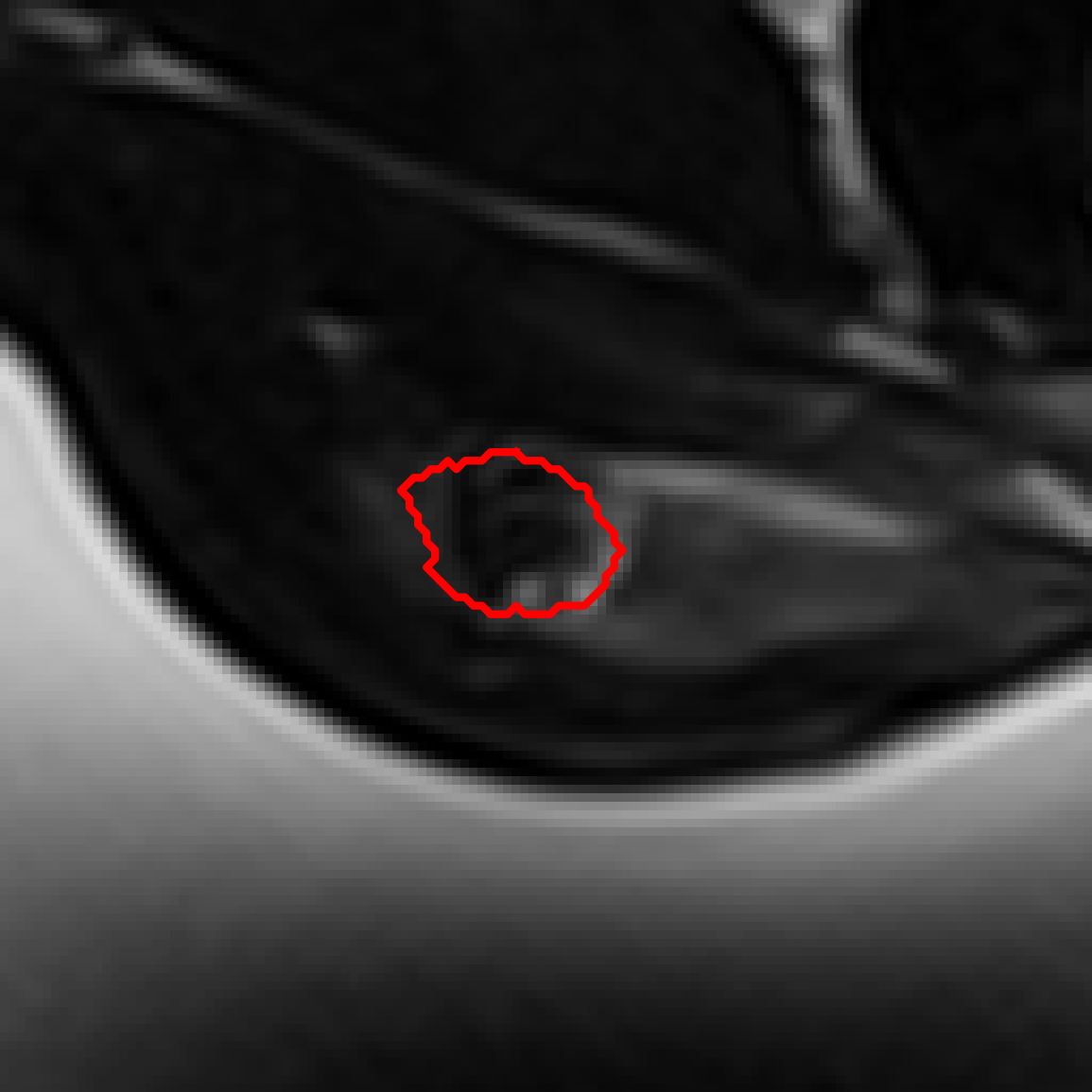}};
        \draw (-1.3,0.5) node[text=white]{\large \textbf{(b)}};
        \draw[white, line width = 2pt] (0.85,-0.6) -- (1.5,-0.6);
        \draw ((1.15,-0.4) node[text=white]{6mm};
        \end{tikzpicture} \\ 
    \begin{tikzpicture}
        \draw (0, 0) node[inner sep=0] {\includegraphics[
        width=\mynlength,
        height=\mynheight,
        trim={0.4in 1.0in 0.6in 1.0in},clip]
        {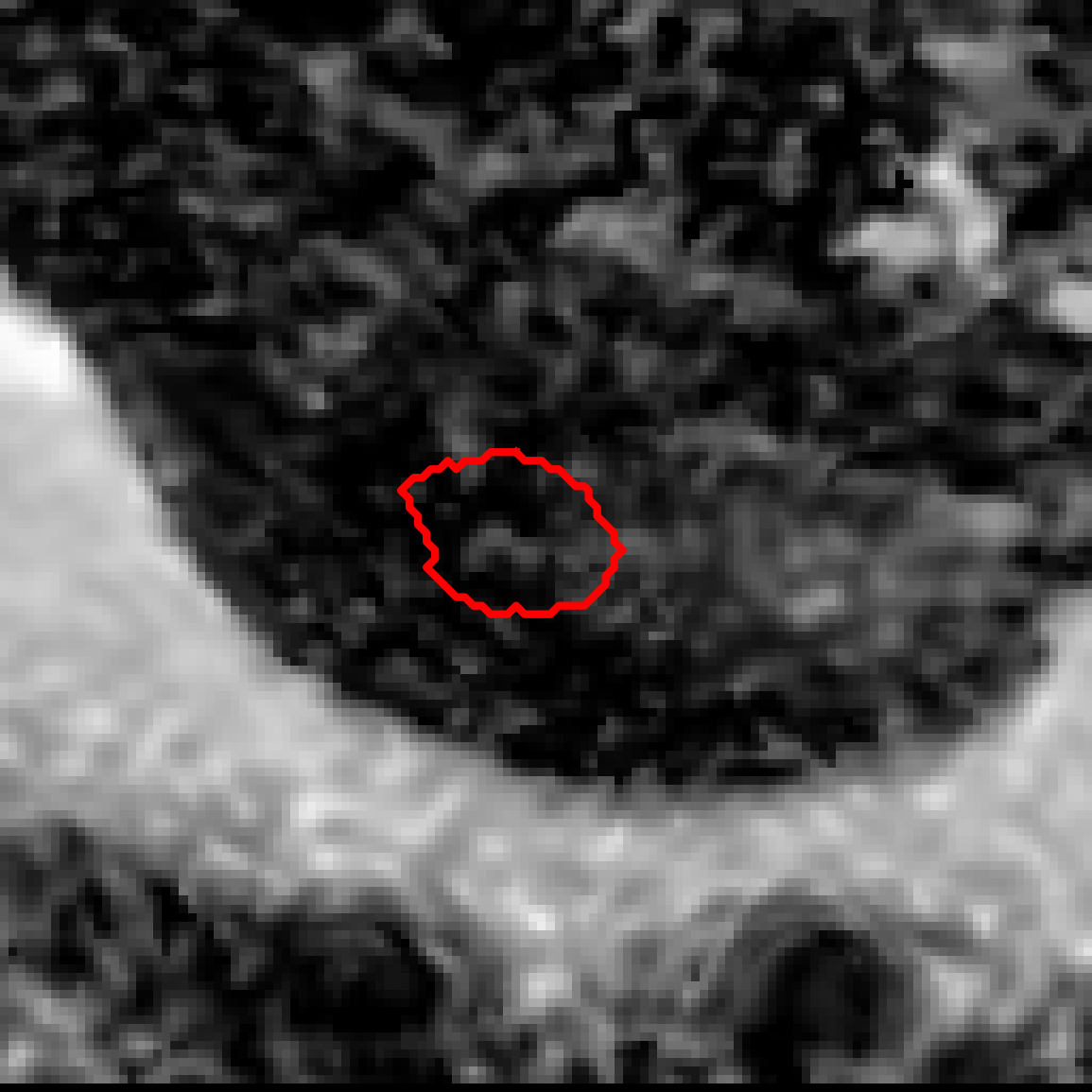}};
        \draw (-1.3,0.5) node[text=white]{\large \textbf{(c)}};
        \draw[white, line width = 2pt] (0.85,-0.6) -- (1.5,-0.6);
        \draw ((1.15,-0.4) node[text=white]{6mm};
        \end{tikzpicture} &
     \begin{tikzpicture}
        \draw (0, 0) node[inner sep=0] {\includegraphics[
        width=\mynlength,
        height=\mynheight,
        trim={0.4in 1.0in 0.6in 1.0in},clip]
        {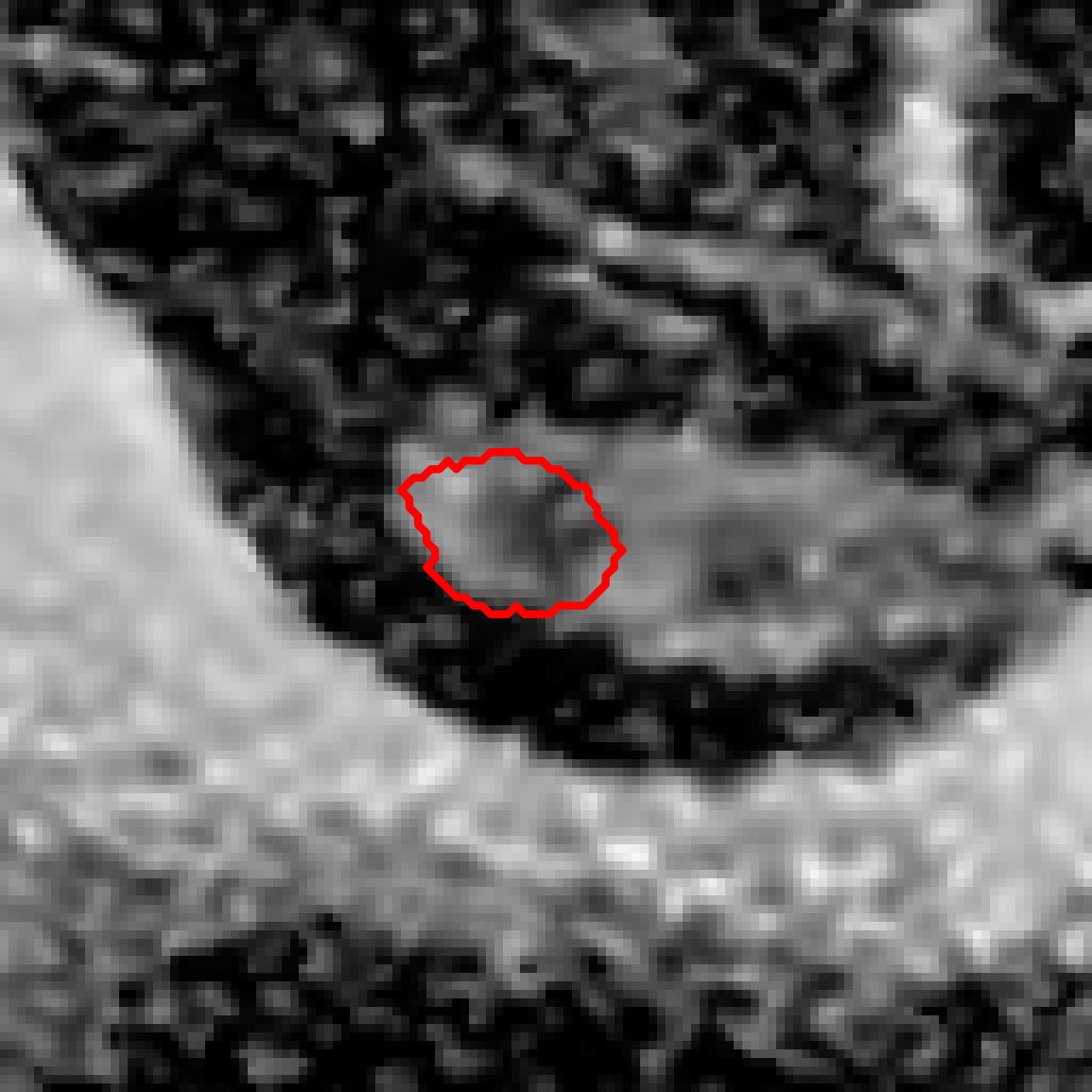}};
        \draw (-1.3,0.5) node[text=white]{\large \textbf{(d)}};
        \draw[white, line width = 2pt] (0.85,-0.6) -- (1.5,-0.6);
        \draw ((1.15,-0.4) node[text=white]{6mm};
        \end{tikzpicture} \\ 
     \begin{tikzpicture}
        \draw (0, 0) node[inner sep=0] {\includegraphics[
        width=\mynlength,
        height=\mynheight,
        trim={0.4in 1.0in 0.6in 1.0in},clip]
        {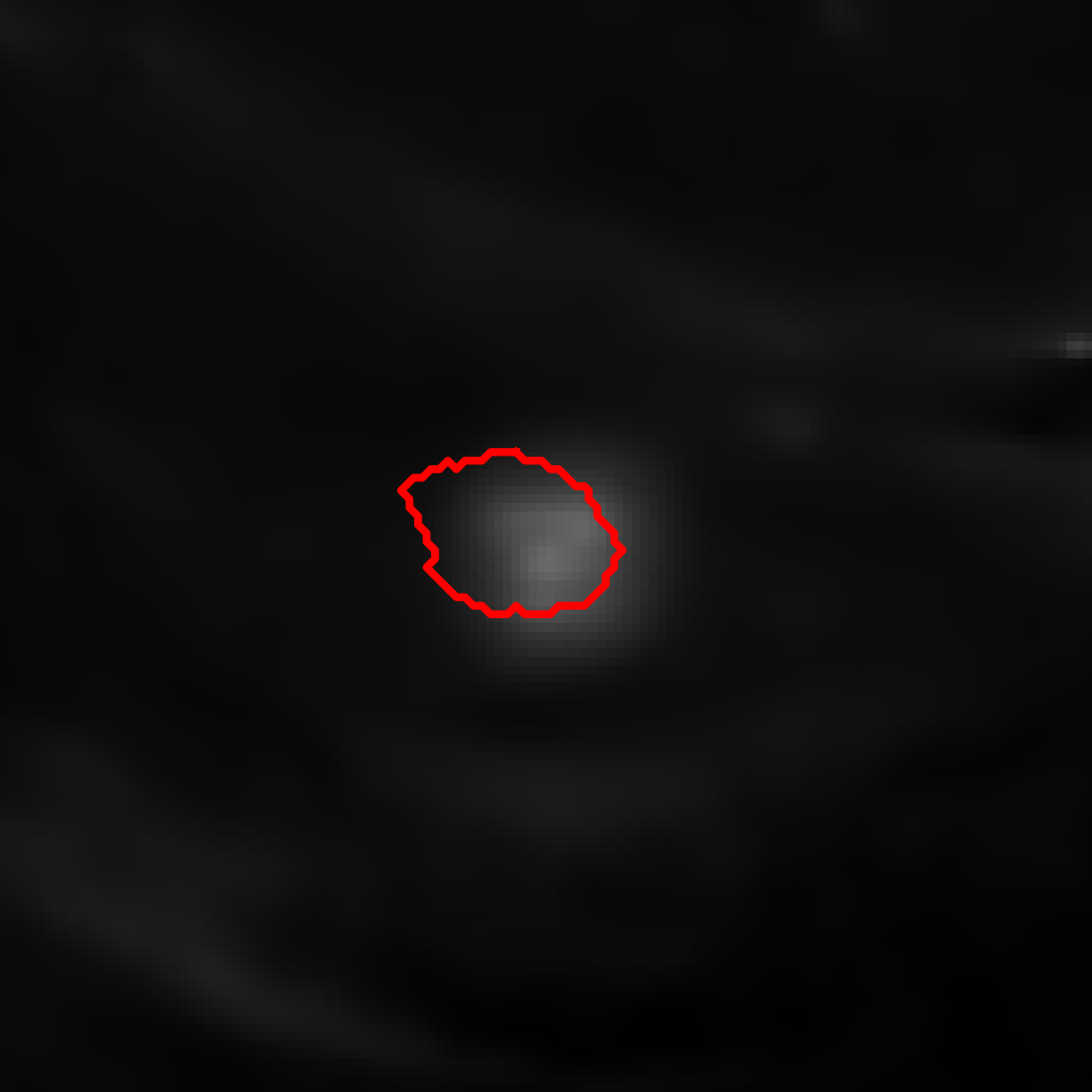}};
        \draw (-1.3,0.5) node[text=white]{\large \textbf{(e)}};
        \draw[white, line width = 2pt] (0.85,-0.6) -- (1.5,-0.6);
        \draw ((1.15,-0.4) node[text=white]{6mm};
        \end{tikzpicture} & 
     \begin{tikzpicture}
        \draw (0, 0) node[inner sep=0] {\includegraphics[
        width=\mynlength,
        height=\mynheight,
        trim={0.4in 1.0in 0.6in 1.0in},clip]
        {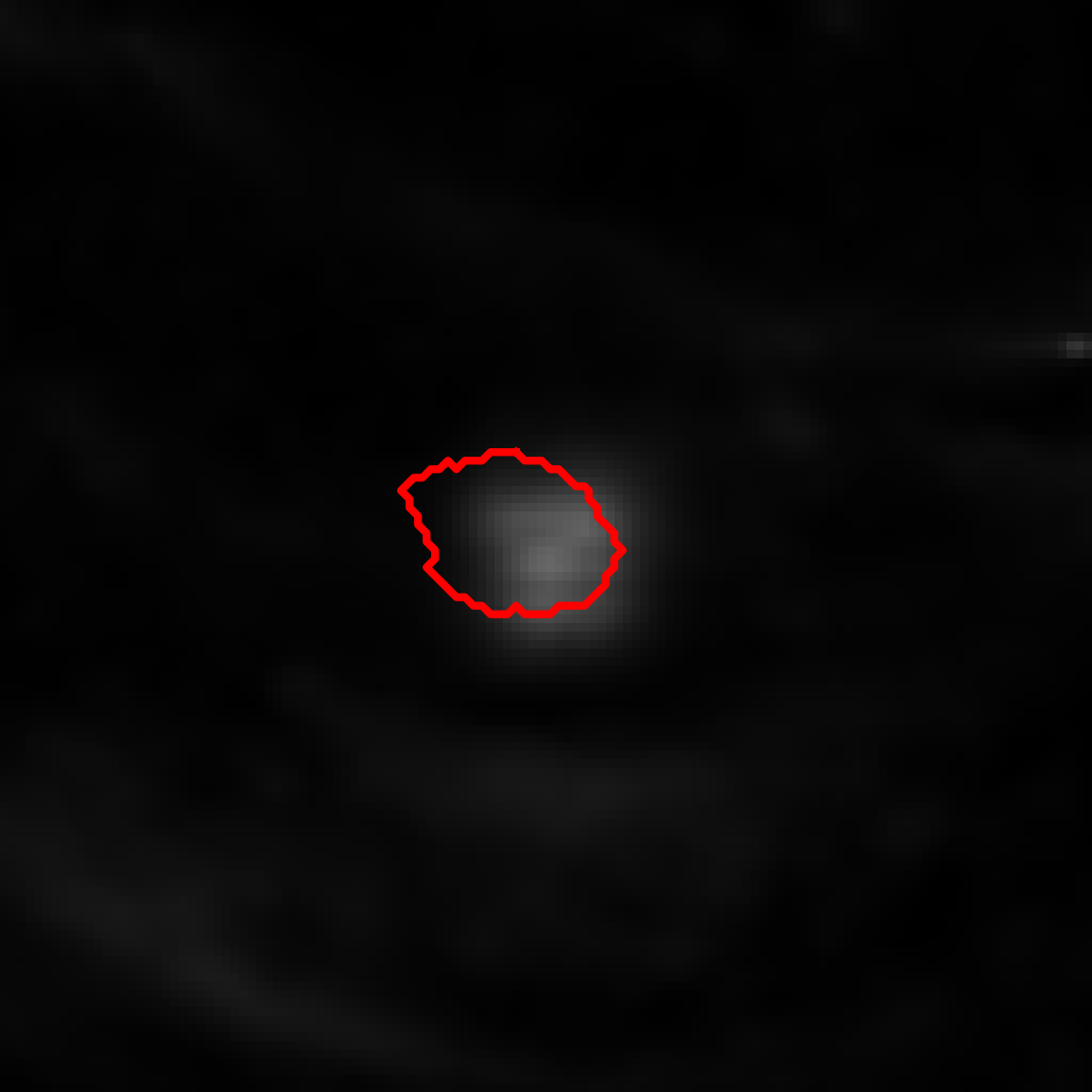}};
        \draw (-1.3,0.5) node[text=white]{\large \textbf{(f)}};
        \draw[white, line width = 2pt] (0.85,-0.6) -- (1.5,-0.6);
        \draw ((1.15,-0.4) node[text=white]{6mm}; 
        \end{tikzpicture} \\
    \multicolumn{2}{c}{\begin{tikzpicture}
        \draw (0, 0) node[inner sep=0] {\includegraphics[
        width=0.55\columnwidth,
        height=0.8in,
        trim={0.4in 1.0in 0.6in 1.0in},clip]
        {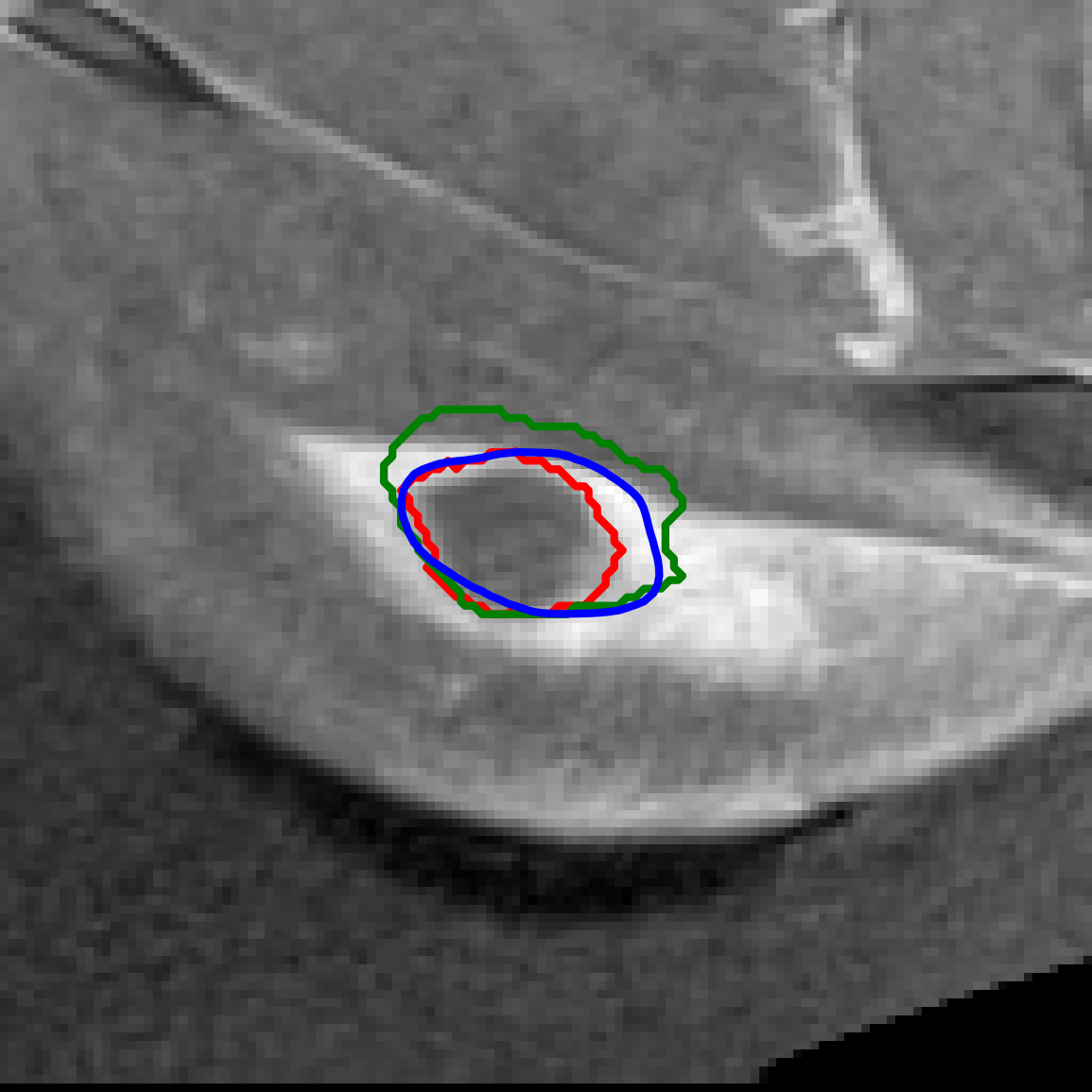}};
        \draw (-1.9,0.7) node[text=white]{\large \textbf{(g)}};
        \draw[white, line width = 2pt] (1.35,-0.7) -- (2.1,-0.7);
        \draw ((1.725,-0.5) node[text=white]{6mm};        
        \end{tikzpicture}}
\end{tabular}
\end{minipage}
\caption{CNN prediction example for subject 7. Normalized input parameters to the network include: (a) pre-T2w, (b) post-T2w, (c) pre-ADC map, (d) post-ADC map, (e) CTD map, and (f) MTP map. The red line on each input parameter shows the registered follow-up NPV contour. (g) The follow-up NPV contour (red), the acute NPV prediction (green), and the CNN prediction (blue).}
\label{fig:cnn}
\end{figure}
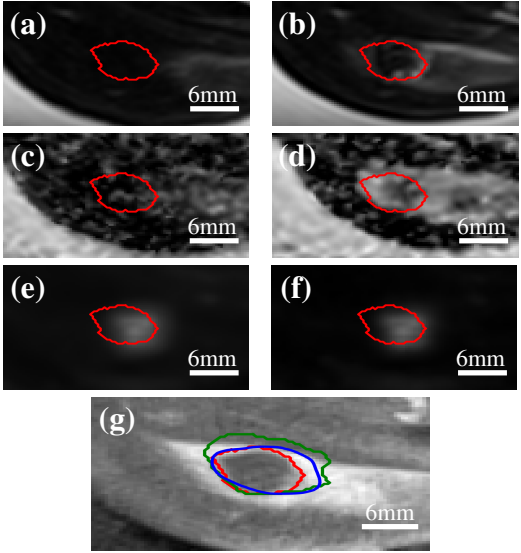 % CNN results
An example of the prediction from the biomarker CNN on one of the validation images is shown in \figureautorefname~\ref{fig:cnn}.
Images (a)-(f) show the six input channels to the network with the registered follow-up NPV contour overlaid in red. 
The bottom image (g) shows the prediction using acute NPV (green contour) versus using the trained biomarker CNN (blue contour) against the follow-up NPV (red contour). 
The network yields a more accurate prediction of the final treated tissue than the acute NPV.
The DICE coefficient of the acute NPV prediction compared with the follow-up NPV over all validation images ranged from 0.00 to 0.83 (0.53 $\pm$ 0.30), whereas the DICE coefficient of the MPB-CNN prediction compared with the follow-up NPV over all validation images ranged from  0.30 to 0.86 (0.71 $\pm$ 0.21).
The DICE coefficent comparisons on a subject-by-subject basis are shown in \tableautorefname~\ref{tab:dice_comp}. 

%% file: discussion.tex
\begin{table}[!t]
  \ra{1.1}
  \centering
  \caption{DICE comparison between the MPB-CNN biomarker and the Acute NPV biomarker for each subject (subjects 2 and 5 were excluded). Each biomarker was compared against the Day 3-5 NPV label. The best DICE score for each subject is bold.}
  \small
  \begin{tabular}{cccccccc}
    \toprule
    & \multicolumn{6}{c}{Subject} \\
    \cmidrule(lr){2-7}
  Marker & 1 & 3 & 4 & 6 & 7 & 8 & Mean $\pm$ Std.\\
    \midrule
    Acute NPV & 0.83 & \textbf{0.71} & 0.00 & 0.23 & 0.70 & 0.70 & 0.53 $\pm$ 0.30\\ 
    MPB-CNN & \textbf{0.86} & 0.65 & \textbf{0.30} & \textbf{0.80} & \textbf{0.86} & \textbf{0.77} & \textbf{0.71 $\pm$ 0.21}\\ 
    \bottomrule
  \end{tabular}
  \label{tab:dice_comp}
\end{table} % Volume table
\section{Discussion}

The results of MPB-CNN show that noncontrast, multiparametric MR biomarkers acquired during the treatment are more reliable for assessing MRgFUS treated tissue than the clinical NPV biomarker.
The MPB-CNN was able to better predict the treated tissue seen on follow-up imaging than the the clinical treatment-day NPV biomarker for five out of six subjects. 
The prediction was based on MR images that do not require gadolinium contrast agents.
This allows treatment to continue following assessment without the risk of trapping toxic contrast agent in the tissue. 
Additionally, the total MR acquisition time of the MPMR images was only \textasciitilde5.5 minutes, which is well within expected time limits for an assessment protocol \cite{futureneedsHectors2016}. 

The MPB-CNN prediction could be improved with additional MR parameters such as quantitative T1 or T2 mapping. The intrinsic T1 and T2 relaxation times are influenced by HIFU-induced changes that could correlate with tissue viability. Several studies have reported significant changes in T1, T2, and ADC values after thermal ablation \cite{futureneedsHectors2016}. These properties are sensitive to post-treatment tissue changes such as hemorrhage, edema, intra- and extra-cellular water content, which may correlate with tissue viability \cite{futureneedsHectors2016}. Although T1- and T2-weighted images are sensitive to these changes in relaxation times, their signal intensities are only relatively quantifiable and dependent on magnet strength, vendor, B0 drift, and scanning parameters. In contrast, quantitative T1 and T2 mapping provide intrinsic tissue properties independent of these factors. Therefore, quantitative inputs to the MPB-CNN may be more easily generalizable across patients or even studies. There is an ongoing, considerable effort in the field of MR to improve standardization and acquisition time of quantitative maps. Improved MR sequences are being developed to simultaneously acquire these MR contrasts to eliminate the least predictive input scans and optimize scan time \cite{cheng2016dual}.

General-purpose registration methods, such as DRAMMS, can cause volume change during registration, as shown in \tableautorefname~\ref{tab:volume_change}. 
Even though the ablation volume was not visible in the source and target registration volumes, DRAMMS still introduced volume changes to these regions.
The average volume change over the 8 subjects using DRAMMS was 21.48\%.
These volume changes are not plausible and are an artifact of the DRAMMS registration algorithm. 
If registration introduces a volume artifact, it will directly affect the prediction of the treatment outcome.
If the volume of the follow-up NPV is increased or decreased during registration, the MPB-CNN would be trained to similarly over- or underestimate the nonviable tissue. 
AVOCADO constrains volume change so the original volume of the follow-up NPV did not change after applying the diffeomorphism.  
This deformation model is more physiologically relevant because tissues do not change volume during deformation. 
In addition to preserving the volume, our registration method out-performed registration accuracy of state-of-the-art DRAMMS registration method for this task. 

Longitudinal registration of MRgFUS images still presents several challenges. 
The water used for acoustic coupling to the quadriceps created a large homogeneous intensity on the MR images, which causes automatic affine registration to favor matching the water signal as opposed to the quadriceps.
Additionally, due to the rabbit being positioned on its side, the delineation between bladder and the quadriceps was often unclear, which made registration of just the quadriceps difficult.
Although we encountered several registration challenges with the chosen model, MRgFUS ablation therapies are being investigated in application to other anatomies, including breast, brain, and prostate.
With other anatomies, challenges specific to this study (such as the delineation between bladder and quadriceps) will not affect the registration; however, similar anatomy-specific issues may require individual attention.
For example, in abdominal applications, changes in bladder fluid levels and bowel contents may induce volume changes in the images. Using patient-specific segmentations, AVOCADO can be easily extended to have a spatially varying volume preservation using the segmentations to define $\alpha(x)$ in \equationautorefname~\ref{eqn:spatial_vary} to not constrain regions of the image undergoing known volume change, such as the fluid in the bladder. This would allow volume preserving registration in the ablated tissue regions between images with large volume changes due to contents of the bowel or bladder.

Despite these limitations, AVOCADO demonstrated that registration with volume conservation does not inhibit the accuracy of registration.
Conversely, AVOCADO out performed DRAMMS because it produced deformations that are more indicative of physiological changes in patient position and pose.
Whereas this study was implemented on a MRgFUS data set, the registration pipeline is generalizable to other longitudinal imaging studies with both therapeutic and diagnostic implications where volume preservation may be needed. 

Although this initial MPB-CNN study presents promising results, it still had limitations. 
We recognize that the network presented here was trained using limited data.
Obtaining data for a single subject to train the network is cumbersome and expensive. 
With limited retrospective data, the focus was on developing a promising network. 
However, future studies need to utilise the MPB-CNN to predict tissue viability following an ablation procedure and compare against the NPV biomarker 3-5 days later to fully test the efficacy of the network on subjects independent of training subjects.
Additional data would allow for training more accurate and generalizable models.
However, this work demonstrates that a CNN approach can provide more accurate measures of the treated tissue compared to traditional methods.
Comparing immediate MPMR biomarkers with the follow-up NPV is a necessary intermediate step toward defining new immediate MPMR biomarkers for treatment assessment. 

Although the MPB-CNN performed better than the NPV biomarker in predicting tissue necrosis, the performance must still be improved to be a accurate biomarker for malignant tumors.
Future work will include adding additional input features to improve the prediction accuracy and subsequently analyzing the MPB-CNN to determine the inputs that provide the most predictive power for the network. 
With a better understanding of the optimal input features, the network could be retrained to provide a more accurate prediction of the treated tissue. 
By understanding the network weights, the underlying intrinsic tissue property that is the best indicator of tissue viability could be determined. 
Gaining this insight will allow the design of a more targeted MR protocol and modify the network inputs to maximize the predictive power of the MPB-CNN. 
Although the presented registration has been applied to a specific animal model and data, the methods can be expanded to investigate and improve other minimally and noninvasive MR guided treatments.

%% file: acknowledgements.tex
\section*{Acknowledgements}
We would like to acknowledge Robb Merrill, Hailey McLean, and Elaine Hillas for their contributions.

%% file: main.bbl
\begin{thebibliography}{10}
\providecommand{\url}[1]{#1}
\csname url@samestyle\endcsname
\providecommand{\newblock}{\relax}
\providecommand{\bibinfo}[2]{#2}
\providecommand{\BIBentrySTDinterwordspacing}{\spaceskip=0pt\relax}
\providecommand{\BIBentryALTinterwordstretchfactor}{4}
\providecommand{\BIBentryALTinterwordspacing}{\spaceskip=\fontdimen2\font plus
\BIBentryALTinterwordstretchfactor\fontdimen3\font minus
  \fontdimen4\font\relax}
\providecommand{\BIBforeignlanguage}[2]{{%
\expandafter\ifx\csname l@#1\endcsname\relax
\typeout{** WARNING: IEEEtran.bst: No hyphenation pattern has been}%
\typeout{** loaded for the language `#1'. Using the pattern for}%
\typeout{** the default language instead.}%
\else
\language=\csname l@#1\endcsname
\fi
#2}}
\providecommand{\BIBdecl}{\relax}
\BIBdecl

\bibitem{haider2008dynamic}
M.~A. Haider \emph{et~al.}, ``Dynamic contrast-enhanced magnetic resonance
  imaging for localization of recurrent prostate cancer after external beam
  radiotherapy,'' \emph{International Journal of Radiation Oncology* Biology*
  Physics}, vol.~70, no.~2, pp. 425--430, 2008.

\bibitem{rouviere2012prostate}
O.~Rouvi{\`e}re, A.~Gelet, S.~Crouzet, and J.-Y. Chapelon, ``Prostate focused
  ultrasound focal therapy—imaging for the future,'' \emph{Nature Reviews
  Clinical Oncology}, vol.~9, no.~12, p. 721, 2012.

\bibitem{kirkham2008mr}
A.~P. Kirkham, M.~Emberton, I.~M. Hoh, R.~O. Illing, A.~A. Freeman, and
  C.~Allen, ``Mr imaging of prostate after treatment with high-intensity
  focused ultrasound,'' \emph{Radiology}, vol. 246, no.~3, pp. 833--844, 2008.

\bibitem{stewart2003focused}
E.~A. Stewart \emph{et~al.}, ``Focused ultrasound treatment of uterine fibroid
  tumors: safety and feasibility of a noninvasive thermoablative technique,''
  \emph{American journal of obstetrics and gynecology}, vol. 189, no.~1, pp.
  48--54, 2003.

\bibitem{tempany2003mr}
C.~M. Tempany, E.~A. Stewart, N.~McDannold, B.~J. Quade, F.~A. Jolesz, and
  K.~Hynynen, ``Mr imaging--guided focused ultrasound surgery of uterine
  leiomyomas: a feasibility study,'' \emph{Radiology}, vol. 226, no.~3, pp.
  897--905, 2003.

\bibitem{hesley2013mr}
G.~K. Hesley, K.~R. Gorny, and D.~A. Woodrum, ``Mr-guided focused ultrasound
  for the treatment of uterine fibroids,'' \emph{Cardiovascular and
  interventional radiology}, vol.~36, no.~1, pp. 5--13, 2013.

\bibitem{gianfelice2008palliative}
D.~Gianfelice, C.~Gupta, W.~Kucharczyk, P.~Bret, D.~Havill, and M.~Clemons,
  ``Palliative treatment of painful bone metastases with mr imaging--guided
  focused ultrasound,'' \emph{Radiology}, vol. 249, no.~1, pp. 355--363, 2008.

\bibitem{zaccagna2015palliative}
F.~Zaccagna \emph{et~al.}, ``Palliative treatment of painful bone metastases
  with mr imaging--guided focused ultrasound surgery: a two-centre study,''
  \emph{Journal of therapeutic ultrasound}, vol.~3, no.~1, p. O51, 2015.

\bibitem{leslie2008high}
T.~Leslie \emph{et~al.}, ``High-intensity focused ultrasound ablation of liver
  tumours: can radiological assessment predict the histological response?''
  \emph{The British journal of radiology}, vol.~81, no. 967, pp. 564--571,
  2008.

\bibitem{wijlemans2012magnetic}
J.~Wijlemans \emph{et~al.}, ``Magnetic resonance-guided high-intensity focused
  ultrasound (mr-hifu) ablation of liver tumours,'' \emph{Cancer Imaging},
  vol.~12, no.~2, p. 387, 2012.

\bibitem{sabel2004cryoablation}
M.~S. Sabel \emph{et~al.}, ``Cryoablation of early-stage breast cancer:
  work-in-progress report of a multi-institutional trial,'' \emph{Annals of
  surgical oncology}, vol.~11, no.~5, pp. 542--549, 2004.

\bibitem{futureneedsHectors2016}
S.~J. Hectors, I.~Jacobs, C.~T. Moonen, G.~J. Strijkers, and K.~Nicolay, ``Mri
  methods for the evaluation of high intensity focused ultrasound tumor
  treatment: Current status and future needs,'' \emph{Magnetic resonance in
  medicine}, vol.~75, no.~1, pp. 302--317, 2016.

\bibitem{hectors2014multiparametric}
S.~J. Hectors, I.~Jacobs, G.~J. Strijkers, and K.~Nicolay, ``Multiparametric
  mri analysis for the identification of high intensity focused
  ultrasound-treated tumor tissue,'' \emph{PloS one}, vol.~9, no.~6, p. e99936,
  2014.

\bibitem{payne2013vivo}
A.~Payne \emph{et~al.}, ``In vivo evaluation of a breast-specific magnetic
  resonance guided focused ultrasound system in a goat udder model,''
  \emph{Medical physics}, vol.~40, no.~7, 2013.

\bibitem{wijlemans2013evolution}
J.~W. Wijlemans \emph{et~al.}, ``Evolution of the ablation region after
  magnetic resonance--guided high-intensity focused ultrasound ablation in a
  vx2 tumor model,'' \emph{Investigative radiology}, vol.~48, no.~6, pp.
  381--386, 2013.

\bibitem{mcdannold2006uterine}
N.~McDannold \emph{et~al.}, ``Uterine leiomyomas: Mr imaging--based thermometry
  and thermal dosimetry during focused ultrasound thermal ablation,''
  \emph{Radiology}, vol. 240, no.~1, pp. 263--272, 2006.

\bibitem{hijnen2013stability}
N.~M. Hijnen, A.~Elevelt, and H.~Gr{\"u}ll, ``Stability and trapping of
  magnetic resonance imaging contrast agents during high-intensity focused
  ultrasound ablation therapy,'' \emph{Investigative radiology}, vol.~48,
  no.~7, pp. 517--524, 2013.

\bibitem{hijnen2013magnetic}
N.~M. Hijnen, A.~Elevelt, J.~Pikkemaat, C.~Bos, L.~W. Bartels, and
  H.~Gr{\"u}ll, ``The magnetic susceptibility effect of gadolinium-based
  contrast agents on prfs-based mr thermometry during thermal interventions,''
  \emph{Journal of therapeutic ultrasound}, vol.~1, no.~1, p.~8, 2013.

\bibitem{plata2015feasibility}
J.~C. Plata \emph{et~al.}, ``A feasibility study on monitoring the evolution of
  apparent diffusion coefficient decrease during thermal ablation,''
  \emph{Medical physics}, vol.~42, no.~9, pp. 5130--5137, 2015.

\bibitem{mannelli2009assessment}
L.~Mannelli, S.~Kim, C.~H. Hajdu, J.~S. Babb, T.~W. Clark, and B.~Taouli,
  ``Assessment of tumor necrosis of hepatocellular carcinoma after
  chemoembolization: diffusion-weighted and contrast-enhanced mri with
  histopathologic correlation of the explanted liver,'' \emph{American Journal
  of Roentgenology}, vol. 193, no.~4, pp. 1044--1052, 2009.

\bibitem{wu2006registration}
Q.~Wu, G.~J. Whitman, D.~S. Fussell, and M.~K. Markey, ``Registration of dce mr
  images for computer-aided diagnosis of breast cancer,'' in \emph{2006
  Fortieth Asilomar Conference on Signals, Systems and Computers}.\hskip 1em
  plus 0.5em minus 0.4em\relax IEEE, 2006, pp. 826--830.

\bibitem{ou2011dramms}
Y.~Ou, A.~Sotiras, N.~Paragios, and C.~Davatzikos, ``Dramms: Deformable
  registration via attribute matching and mutual-saliency weighting,''
  \emph{Medical image analysis}, vol.~15, no.~4, pp. 622--639, 2011.

\bibitem{li2009nonrigid}
X.~Li \emph{et~al.}, ``A nonrigid registration algorithm for longitudinal
  breast mr images and the analysis of breast tumor response,'' \emph{Magnetic
  resonance imaging}, vol.~27, no.~9, pp. 1258--1270, 2009.

\bibitem{brock2006feasibility}
K.~K. Brock, L.~A. Dawson, M.~B. Sharpe, D.~J. Moseley, and D.~A. Jaffray,
  ``Feasibility of a novel deformable image registration technique to
  facilitate classification, targeting, and monitoring of tumor and normal
  tissue,'' \emph{International Journal of Radiation Oncology* Biology*
  Physics}, vol.~64, no.~4, pp. 1245--1254, 2006.

\bibitem{ou2015deformable}
Y.~Ou \emph{et~al.}, ``Deformable registration for quantifying longitudinal
  tumor changes during neoadjuvant chemotherapy,'' \emph{Magnetic resonance in
  medicine}, vol.~73, no.~6, pp. 2343--2356, 2015.

\bibitem{jahani2018deformable}
N.~Jahani \emph{et~al.}, ``Deformable image registration as a tool to improve
  survival prediction after neoadjuvant chemotherapy for breast cancer: results
  from the acrin 6657/i-spy-1 trial,'' in \emph{Medical Imaging 2018:
  Computer-Aided Diagnosis}, vol. 10575.\hskip 1em plus 0.5em minus 0.4em\relax
  International Society for Optics and Photonics, 2018, p. 105752S.

\bibitem{humphrey2003continuum}
J.~D. Humphrey, ``Continuum biomechanics of soft biological tissues,''
  \emph{Proceedings of the Royal Society of London. Series A: Mathematical,
  Physical and Engineering Sciences}, vol. 459, no. 2029, pp. 3--46, 2003.

\bibitem{godwin2009healing}
B.~L. Godwin and J.~E. Coad, ``Healing responses following cryothermic and
  hyperthermic tissue ablation,'' in \emph{Energy-based Treatment of Tissue and
  Assessment V}, vol. 7181.\hskip 1em plus 0.5em minus 0.4em\relax
  International Society for Optics and Photonics, 2009, p. 718103.

\bibitem{rohlfing2003volume}
T.~Rohlfing, C.~R. Maurer, D.~A. Bluemke, and M.~A. Jacobs, ``Volume-preserving
  nonrigid registration of mr breast images using free-form deformation with an
  incompressibility constraint,'' \emph{IEEE transactions on medical imaging},
  vol.~22, no.~6, pp. 730--741, 2003.

\bibitem{palussiere2003feasibility}
J.~Palussiere \emph{et~al.}, ``Feasibility of mr-guided focused ultrasound with
  real-time temperature mapping and continuous sonication for ablation of vx2
  carcinoma in rabbit thigh,'' \emph{Magnetic Resonance in Medicine: An
  Official Journal of the International Society for Magnetic Resonance in
  Medicine}, vol.~49, no.~1, pp. 89--98, 2003.

\bibitem{christensen1996deformable}
G.~E. Christensen, R.~D. Rabbitt, and M.~I. Miller, ``Deformable templates
  using large deformation kinematics,'' \emph{IEEE transactions on image
  processing}, vol.~5, no.~10, pp. 1435--1447, 1996.

\bibitem{joshi2000landmark}
S.~C. Joshi and M.~I. Miller, ``Landmark matching via large deformation
  diffeomorphisms,'' \emph{IEEE transactions on image processing}, vol.~9,
  no.~8, pp. 1357--1370, 2000.

\bibitem{guo2006multi}
Y.~Guo and C.-C. Lu, ``Multi-modality image registration using mutual
  information based on gradient vector flow,'' in \emph{18th International
  Conference on Pattern Recognition (ICPR'06)}, vol.~3.\hskip 1em plus 0.5em
  minus 0.4em\relax IEEE, 2006, pp. 697--700.

\bibitem{haker2004optimal}
S.~Haker, L.~Zhu, A.~Tannenbaum, and S.~Angenent, ``Optimal mass transport for
  registration and warping,'' \emph{International Journal of computer vision},
  vol.~60, no.~3, pp. 225--240, 2004.

\bibitem{bauer2014overview}
M.~Bauer, M.~Bruveris, and P.~W. Michor, ``Overview of the geometries of shape
  spaces and diffeomorphism groups,'' \emph{Journal of Mathematical Imaging and
  Vision}, vol.~50, no. 1-2, pp. 60--97, 2014.

\bibitem{younes2010shapes}
L.~Younes, \emph{Shapes and diffeomorphisms}.\hskip 1em plus 0.5em minus
  0.4em\relax Springer, 2010, vol. 171.

\bibitem{misiolek1993stability}
G.~Misio{\l}ek, ``Stability of flows of ideal fluids and the geometry of the
  group of diffeomorphisms,'' \emph{Indiana University mathematics journal},
  pp. 215--235, 1993.

\bibitem{hinkle20124d}
J.~Hinkle, M.~Szegedi, B.~Wang, B.~Salter, and S.~Joshi, ``4d ct image
  reconstruction with diffeomorphic motion model,'' \emph{Medical image
  analysis}, vol.~16, no.~6, pp. 1307--1316, 2012.

\bibitem{bhatia2012helmholtz}
H.~Bhatia, G.~Norgard, V.~Pascucci, and P.-T. Bremer, ``The helmholtz-hodge
  decomposition—a survey,'' \emph{IEEE Transactions on visualization and
  computer graphics}, vol.~19, no.~8, pp. 1386--1404, 2012.

\bibitem{cantarella2002vector}
J.~Cantarella, D.~DeTurck, and H.~Gluck, ``Vector calculus and the topology of
  domains in 3-space,'' \emph{The American mathematical monthly}, vol. 109,
  no.~5, pp. 409--442, 2002.

\bibitem{van1983matrix}
C.~F. Van~Loan and G.~H. Golub, \emph{Matrix computations}.\hskip 1em plus
  0.5em minus 0.4em\relax Johns Hopkins University Press, 1983.

\bibitem{buhmann2003radial}
M.~D. Buhmann, \emph{Radial basis functions: theory and implementations}.\hskip
  1em plus 0.5em minus 0.4em\relax Cambridge university press, 2003, vol.~12.

\bibitem{beg2005computing}
M.~F. Beg, M.~I. Miller, A.~Trouv{\'e}, and L.~Younes, ``Computing large
  deformation metric mappings via geodesic flows of diffeomorphisms,''
  \emph{International journal of computer vision}, vol.~61, no.~2, pp.
  139--157, 2005.

\bibitem{fedorov20123d}
A.~Fedorov \emph{et~al.}, ``3d slicer as an image computing platform for the
  quantitative imaging network,'' \emph{Magnetic resonance imaging}, vol.~30,
  no.~9, pp. 1323--1341, 2012.

\bibitem{ou2014comparative}
Y.~Ou, H.~Akbari, M.~Bilello, X.~Da, and C.~Davatzikos, ``Comparative
  evaluation of registration algorithms in different brain databases with
  varying difficulty: results and insights,'' \emph{IEEE transactions on
  medical imaging}, vol.~33, no.~10, pp. 2039--2065, 2014.

\bibitem{ou2012validation}
Y.~Ou, D.~H. Ye, K.~M. Pohl, and C.~Davatzikos, ``Validation of dramms among 12
  popular methods in cross-subject cardiac mri registration,'' in
  \emph{International Workshop on Biomedical Image Registration}.\hskip 1em
  plus 0.5em minus 0.4em\relax Springer, 2012, pp. 209--219.

\bibitem{dice1945measures}
L.~R. Dice, ``Measures of the amount of ecologic association between species,''
  \emph{Ecology}, vol.~26, no.~3, pp. 297--302, 1945.

\bibitem{van2013cem43}
G.~C. Van~Rhoon, T.~Samaras, P.~S. Yarmolenko, M.~W. Dewhirst, E.~Neufeld, and
  N.~Kuster, ``Cem43° c thermal dose thresholds: a potential guide for
  magnetic resonance radiofrequency exposure levels?'' \emph{European
  radiology}, vol.~23, no.~8, pp. 2215--2227, 2013.

\bibitem{dewhirst2003basic}
M.~W. Dewhirst, B.~Viglianti, M.~Lora-Michiels, M.~Hanson, and P.~Hoopes,
  ``Basic principles of thermal dosimetry and thermal thresholds for tissue
  damage from hyperthermia,'' \emph{International journal of hyperthermia},
  vol.~19, no.~3, pp. 267--294, 2003.

\bibitem{ronneberger2015u}
O.~Ronneberger, P.~Fischer, and T.~Brox, ``U-net: Convolutional networks for
  biomedical image segmentation,'' in \emph{International Conference on Medical
  image computing and computer-assisted intervention}.\hskip 1em plus 0.5em
  minus 0.4em\relax Springer, 2015, pp. 234--241.

\bibitem{cheng2016dual}
C.-C. Cheng \emph{et~al.}, ``Dual-pathway multi-echo sequence for simultaneous
  frequency and t2 mapping,'' \emph{Journal of Magnetic Resonance}, vol. 265,
  pp. 177--187, 2016.

\end{thebibliography}
